\documentclass[9pt,twocolumn,letterpaper]{IEEEtran}

\usepackage{mathptmx}
\usepackage[scaled=-90]{helvet}
\usepackage{courier}

\usepackage{amsmath}
\usepackage{amsfonts,amssymb}
\usepackage{graphicx}

\usepackage{subfigure}

\newtheorem{theorem}{Theorem}[section]
\newtheorem{conjecture}[theorem]{Conjecture}
\newtheorem{corollary}[theorem]{Corollary}
\newtheorem{proposition}[theorem]{Proposition}
\newtheorem{lemma}[theorem]{Lemma}
\newtheorem{definition}[theorem]{Definition}

\newtheorem{fact}[theorem]{Fact}

\newcommand{\goodgap}{%
\hspace{\subfigtopskip}%
\hspace{\subfigbottomskip}} 

\newcommand{\POPS}{{\mathrm{POPS}}(d,g)}
\newcommand{\POPSg}{{\mathrm{POPS}}(g,g)}
\newcommand{\PR}{{\mathbf{Pr}}}
\newcommand{\E}{{\mathbf{E}}}

\begin{document}

\title{On-Line Permutation Routing\\ in Partitioned Optical Passive Star Networks}

\author{Alessandro Mei\thanks{Alessandro Mei is with the
Department of Computer Science, University of Rome ``La Sapienza'', Italy
(e-mail: mei@di.uniroma1.it).} and Romeo Rizzi\thanks{Romeo Rizzi is with the
Department of Information and Communication Technology,
University of Trento, Italy (e-mail: romeo.rizzi@unitn.it).}}

\maketitle

\begin{abstract}
This paper establishes the state of the art in both deterministic and randomized online
permutation routing in the POPS network.
Indeed, we show that any permutation can be routed online on a $\POPS$ network
either with $O(\frac{d}{g}\log g)$ deterministic slots, or, with high probability,
with $5c\lceil d/g\rceil+o(d/g)+O(\log\log g)$ randomized slots,
where constant~$c=\exp (1+e^{-1})\approx 3.927$.
When $d=\Theta(g)$, that we claim to be the ``interesting'' case, the
randomized algorithm
is exponentially faster than any other algorithm in the literature, both deterministic and randomized
ones. This is true in practice as well. Indeed, experiments show that it
outperforms its rivals even starting
from as small a network as a ${\mathrm{POPS}}(2,2)$, and the gap grows exponentially with the
size of the network. We can also show that, under proper hypothesis,
no deterministic algorithm can asymptotically match its performance.
\end{abstract}


\maketitle

\section{Introduction}

The ever-growing demand of fast interconnections in multiprocessor systems
has fostered a large interest in optical technology. All-optical communication
benefits from a number of good characteristics such as no opto-electronic
conversion, high noise immunity, and low latency. Optical technology can
provide an enormous amount of bandwidth and, most probably, will
have an important role in the future of distributed and parallel computing
systems.

\begin{figure}
\begin{center}
\includegraphics[scale=.7]{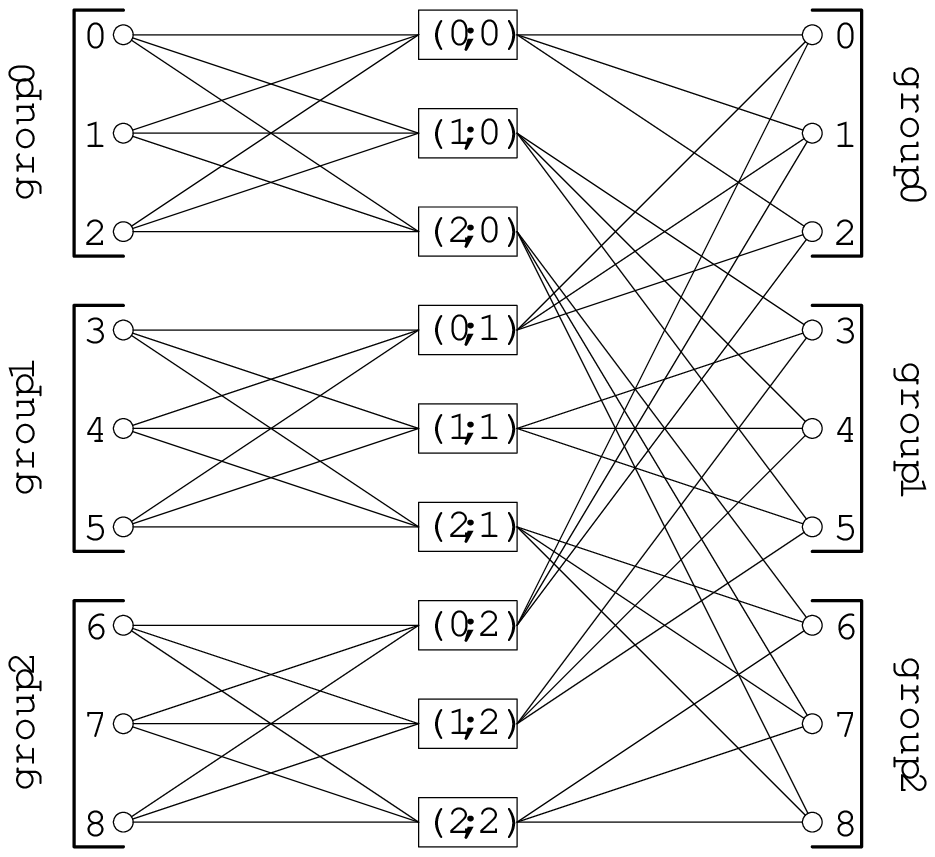}
\end{center}
\caption{A ${\mathrm{POPS}}(3,3)$. Processors are shown as circles, while optical passive stars
are shown as boxes. Optical signals flow from the left to the right.
The processors on the left and the processors on the right are the same
objects shown twice for the sake of clearness.}
\label{fig:pops}
\end{figure}

The Partitioned Optical Passive Stars (POPS)
network~\cite{clmtg94,gmclt95,gm98,mgcl98} is a SIMD
parallel architecture that uses a fast optical network composed of
multiple Optical Passive Star (OPS) couplers. 
A $d\times d$ OPS coupler is
an all-optical passive device which is capable of
receiving an optical signal from one of its $d$ sources and broadcast it to
all of its $d$ destinations.
The number of processors of the network is denoted by $n$, and each processor
has a distinct index in $\{0,\dotsc, n-1\}$.
The $n$ processors are partitioned into $g$
groups of $d$ processors, $n=dg$, in such a way that processor $i$ belongs to
group~${\mathrm{group}}(i):=\lfloor i/d\rfloor$ (see Figure~\ref{fig:pops}).
For each pair of groups $a,b\in\{0,\dotsc,g-1\}$, a coupler~$c(b,a)$ is
introduced which has all the $d$ processors of group~$a$ as sources and all
the $d$ processors of group~$b$ as destinations.
During a computational step (also referred to as a \emph{slot}), each
processor~$i$ receives a single message from one of the $g$ couplers
$c({\mathrm{group}}(i), a)$, $a\in\{0,\dotsc,g-1\}$, performs some
local computations, and sends a single message to a subset of the $g$ couplers
$c(b, {\mathrm{group}}(i))$, $b\in\{0,\dotsc,g-1\}$. The couplers are
broadcast devices, so this message can be received by more than one processor
in the destination groups.
In agreement with the literature, in the case when multiple
messages are sent to the same coupler, we assume that no message is delivered.
This architecture is denoted by $\POPS$.

One of the advantages of a $\mathrm{POPS}(d,g)$ network is
that its diameter is one. A packet can
be sent from processor~$i$ to processor~$j$, $i\neq j$, in one slot
by using coupler $c(\mathrm{group}(j),\mathrm{group}(i))$. However,
its bandwidth varies according to $g$. In a $\mathrm{POPS}(n,1)$ network,
only one packet can be sent through the single coupler per slot.
On the other extreme, a $\mathrm{POPS}(1,n)$ network is a highly expensive,
fully interconnected optical network using $n^2$ OPS couplers.
A one-to-all communication pattern can also be performed in only one slot in
the following way: Processor~$i$ (the speaker) sends the packet to
all the couplers~$c(a,\mathrm{group}(i))$, $a\in\{0,\dotsc,g-1\}$,
during the same slot all the processors~$j$, $j\in\{0,\dotsc,n-1\}$,
can receive the packet through coupler
$c(\mathrm{group}(j),\mathrm{group}(i))$.

The POPS network has been shown to support a number of non trivial
algorithms. Several common communication patterns are realized
in~\cite{gm98}. Simulation algorithms for the ring, the mesh, and the hypercube interconnection
networks can be found in~\cite{gm-MPPUOI95} and~\cite{s00a}. Some reliability issues
are analyzed in~\cite{c-LCN97}. Algorithms for data sum, prefix
sums, consecutive sum, adjacent sum, and several data movement operations
are also described in~\cite{s00a} and~\cite{ds-IEEETPDS03}. Later, both the algorithms
for hypercube simulation and prefix sums have been improved in~\cite{mr-HIPC03}.
An algorithm for matrix
multiplication is provided in~\cite{s00b}. 
Moreover, \cite{bf96} shows that POPS networks can be modeled by directed
and complete stack graphs with loops, and uses this formalization to
obtain optimal embeddings of rings and de Bruijn graphs into POPS
networks.

In~\cite{ds-IEEETPDS03}, Datta and Soundaralakshmi claim that in most practical
$\POPS$ networks it is likely that $d>g$. We believe that they are only partly
right. While it is true that
systems with $d\ll g$ are too expensive, it is also true that systems with $d\gg g$ give
too low parallelism to be worth building. We illustrate our point with an example.
Consider the problem of summing $16n$ data values on a $\POPS$ network,
$d=g=\sqrt{n}$. This network has $n$ processors. Therefore, the algorithm can work as follows:
we input 16 data values per processor, let each processor sum up its 16 data values, and
finally we use the algorithm in~\cite{ds-IEEETPDS03} to get the overall sum. This algorithm
requires 16 steps to input the data values and compute the local sums, plus
$2\log\sqrt{n}=\log n$ slots for computing the final result. A total of $16+\log n$ slots.
With the idea of upgrading our system,
we buy additional $15n$ processors and build a $16n$ processor ${\mathrm{POPS}}(d',g')$ network
with $d'=16d=16\sqrt{n}$ and $g'=g=\sqrt{n}$.
Now, we can use just one step to input the data values, one per processor, and then
use the same algorithm in~\cite{ds-IEEETPDS03} to get the overall sum. Unfortunately,
this algorithm still requires $16+\log n$ slots, even though we are solving a problem of the
same size using a system with 16 times more processors!

The problem is not on the data sum algorithm in~\cite{ds-IEEETPDS03}. Essentially the same thing
happens with the prefix sums algorithm in~\cite{ds-IEEETPDS03}, the simulations in~\cite{s00a},
and all the other algorithms in the literature for the POPS network we know of, including the ones
presented in this paper. The point is that a $\POPS$
network can exchange $g^2$ messages at most in a slot. This is an unavoidable bottleneck
for networks where $d$ is much larger than $g$, resulting in the poor parallelism of
these systems.
Also, experience says that the case $d=g$ is the most interesting from a
``mathematical'' point of view. In the past literature, the case $d>g$ and symmetrically the case $d<g$
are always dealt with by reducing them to the case $d=g$, that usually contains the
``core'' of the problem in its purest form. This work is not an exception to this empirical
yet general rule.
So, it is probably more reasonable to assume that practical POPS networks
will have $d=\Theta(g)$, that is $d/g$,
and similarly $g/d$, bounded by a constant.

In any case, finding good algorithms for the case $d\neq g$, both $d<g$ and
$d>g$, is of absolute
importance, since it is not clear what is the optimal tradeoff between $d$, $g$, and the cost
of the network yet. Furthermore, an optimal tradeoff may not exist in general,
since it probably depends on the specific problem being solved.
By the way, such algorithms are often non trivial, as, for example,
in~\cite{ds-IEEETPDS03}. Therefore, we partly accept the claim in~\cite{ds-IEEETPDS03}
that the number of groups cannot substantially exceed the number of processors per
group. So, throughout the whole paper, we will discuss our asymptotical results assuming
that $g$ grows and that $d=\Omega(g)$. Nonetheless, we will
keep in mind that the ``important'' case is likely to be $d=\Theta(g)$.

Here, we consider the \emph{permutation routing problem}: Each of the $n$ processors of the POPS
network has a packet that is to be sent to another node, and each processor is the destination
of exactly one packet. This is a fundamental problem in parallel computing and
interconnection networks, and the literature on this topic is vast. As an excellent starting point,
the reader can see~\cite{l92}. On the POPS network, this problem has been studied in two
different versions: the \emph{offline} and the \emph{online} permutation routing problem.
In the former, the permutation to be routed is globally known in the network. Therefore,
every processor can pre-compute the route for its packet taking advantage of this information.
This version of the problem has been implicitly studied, for particular permutations, in
all the simulation algorithms we reviewed above. Later, most of these results
have been unified by proving that any permutation can optimally be routed off-line
in one slot, when $d=1$, and $2\lceil d/g\rceil$ slots, when $d>1$~\cite{mr-JPDC03}.

In the online version, every processor knows only the destination of the packet it stores.
This problem has been attacked in~\cite{ds-IEEETPDS03}. The solution 
iteratively makes use of a sub-routine that sorts $g^2$ items in ${\mathrm{POPS}}(g,g)$
subnetworks of the larger $\POPS$ network. The sub-routine is built by hypercube simulation
starting from either Cypher and Plaxton's $O(\log n\log\log n)$ sorting algorithm for the
$n$-processor hypercube or from Leighton's implementation~\cite{l92} on the
$n$-processor hypercube of Batcher's odd-even merge sort algorithm~\cite{b-AFIPS68}. 
In the first case, Datta and Sounderalakshmi get the asymptotically fastest algorithm for
routing in the POPS network, running in $O(\frac{d}{g}\log g\log\log g)$ slots. In the second,
they get an algorithm that turns out to be the fastest in practice, running in
$\frac{8d}{g}\log^2 g+\frac{21d}{g}+3\log g+7$ slots. Recently, and independently
of this work, Rajasekaran and Davila have presented a randomized algorithm for online
permutation routing that runs in $O(\frac{d}{g}+\log g)$ slots~\cite{rd-ICPADS04}.

Our contribution is both theoretical and practical. 
We show that any permutation can be routed on a $\POPS$ network
either with $O(\frac{d}{g}\log g)$ deterministic slots, or, with high probability,
with $5c\lceil d/g\rceil+o(d/g)+O(\log\log g)$ randomized slots,
where constant~$c=\exp (1+e^{-1})\approx 3.927$. The deterministic algorithm
is based on a direct simulation of the AKS network, and it is the first that requires
only $O(\frac{d}{g}\log g)$ slots.
When $d=\Theta(g)$, that we claim to be the ``interesting'' case, the
randomized algorithm
is exponentially faster than any other algorithm in the literature, both deterministic and randomized
ones. This is true in practice as well. Indeed, our experiments show that it
outperforms its rivals even starting
from as small a network as a ${\mathrm{POPS}}(2,2)$, and the gap grows exponentially with the
size of the network. We can also show that, under proper hypothesis,
no deterministic algorithm can asymptotically match its performance.

This paper also presents a strong separation theorem between determinism and randomization.
We build a meaningful and natural problem inspired on permutation routing in the POPS
network such that there exists a $O(\log\log g)$ slots randomized solution, and such that
no deterministic solution can do better than $O(\log g)$ slots, that is exponentially slower.
To the best of our knowledge, this is the first strong separation result from $\log g$ to
$\log\log g$, and, quite interestingly, it does not make use of the notion of oblivious routing,
that we show to be essentially out of target in the context of routing in the POPS network.

\section{A Deterministic Algorithm}

Let ${\mathbb{N}}_m:=\{0,1,\dotsc,m-1\}$ denote the set of the first $m$ natural numbers.
In the \emph{on-line permutation routing problem} we are given $n$ packets, one per
processor. Packet $p_i$, $i\in{\mathbb{N}}_{n}$, originates at processor~$i$, the \emph{source
processor}, and has
processor~$\pi(i)$ as \emph{destination}, where $\pi$ is a permutation of ${\mathbb{N}}_n$.

The problem is to route all the
packets to destination with as few slots as possible.
Crucially, permutation $\pi$ is not known in advance---at the beginning of
the computation, each processor knows only the destination of the packet it stores.

\subsection{The Upper Bound}

So far, the best deterministic algorithm for online permutation routing on the $\POPS$
network is presented in~\cite{ds-IEEETPDS03}. The algorithm runs in $O(\frac{d}{g}\log^2 g)$ slots.
The computational bottleneck is
a $O(\log^2 g)$ sorting sub-routine that sorts $g^2$ data value $\lceil d/g\rceil$ times, each
on one of the $\lceil d/g\rceil$ ${\mathrm{POPS}}(g,g)$
sub-networks into which the larger $\POPS$ network is partitioned. The idea in~\cite{ds-IEEETPDS03}
is to make each ${\mathrm{POPS}}(g,g)$
network simulate Leighton's $O(\log^2 n)$ sorting algorithm for the $n$-processor hypercube~\cite{l92},
that is, in turn, an implementation of Batcher's odd-even merge sort. This is
carried out by using a general result due to Sahni~\cite{s00a},
showing that every move of a \emph{normal}
algorithm for the hypercube (where only one dimension is used for communication at each
step) can be simulated with $2\lceil d/g\rceil$ slots on a POPS network of the same size. Since
Leighton's algorithm is normal, and since
the sub-routine is always used on ${\mathrm{POPS}}(g,g)$ sub-networks, we get a constant
factor slow-down.

The algorithm in~\cite{ds-IEEETPDS03} is fairly good in practice, since hidden constants are small.
However, we are interested in the best asymptotical result. So, as suggested
in~\cite{ds-IEEETPDS03}, we can replace the Leighton implementation of Batcher's odd-even
merge sort with Cypher and Plaxton's routing algorithm for the hypercube,
that is asymptotically faster (though slower for networks of practical size),
since it runs in $O(\log n\log\log n)$ time~\cite{cp-TR90}. This yields
a $O(\frac{d}{g}\log g\log\log g)$ slots algorithm for permutation routing on the POPS network,
that is a good improvement. 
Nonetheless, here we do even better. Our simple key idea
is to simulate a fast sorting network directly on the POPS, instead of going through
hypercube simulation. By giving an improved $O(\log g)$ upper bound for sorting on the POPS network,
we also get an asymptotically faster algorithm for online permutation routing.
 
A \emph{comparator} $[i:j]$, $i,j\in{\mathbb{N}}_n$ sorts the $i$-th and $j$-th element of a data
sequence into non-decreasing order. A \emph{comparator stage} is a composition of comparators
$[i_1:j_1]\circ\dotsb\circ [i_k:j_k]$ such that all $i_r$ and $j_s$ are distinct, and a
\emph{sorting network} is a sequence of comparator stages such that any input sequence
of $n$ data elements is sorted into non-decreasing order.
An introduction to sorting networks can be found
in~\cite{clr92}. Crucially, we can show that a $\POPS$ network can efficiently simulate
any comparator stage.
\begin{theorem}[\cite{mr-JPDC03}]
\label{thm:permutationrouting}
A $\POPS$ network can route off-line any permutation among the $n=dg$
processors using one slot when $d=1$ and $2\lceil d/g\rceil$ slots when $d>1$.
\end{theorem}
\begin{lemma}
\label{lem:comparatorstage}
A  $\POPS$ network, $n=dg$, can simulate a comparator stage in one slot, when $d=1$,
and in $2\lceil d/g\rceil$ slots, when $d>1$.
\end{lemma}
\begin{proof}
Let $[i_1:j_1]\circ\dotsb\circ [i_k:j_k]$ be a comparator stage. We define a function $\pi$ such
that $\pi(i_r)=j_r$ and $\pi(j_r)=i_r$ for all $r$.
Since all $i_r$ are distinct, and so are all $j_s$,
$\pi$ can arbitrarily
be extended in such a way to be a permutation. By
Theorem~\ref{thm:permutationrouting}, $\pi$ can be routed in one slot when $d=1$, and
$2\lceil d/g\rceil$ slots when $d>1$.
During this routing,
for every $r$, processor~$i_r$ sends
its data value to processor~$j_r$ and vice-versa. Then, processor~$i_r$ discards the maximum
of the two data values, while processor~$j_r$ discards the minimum.
\end{proof}
In~\cite{aks-FOCS83}, the AKS sorting network is presented. This network is able to sort any
data sequence with only $O(\log n)$ comparator stages, which is optimal.
By simulating the AKS network
on a POPS network using Lemma~\ref{lem:comparatorstage}, we easily get the following theorem.
\begin{theorem}
\label{thm:deterministico}
A $\POPSg$ network can sort $g^2$ data values in $O(\log g)$ slots.
\end{theorem}
The above result is the key to improve on the best deterministic algorithm for online permutation
routing in the literature.
\begin{corollary}
\label{cor:deterministico}
A $\POPS$ network can route on-line any permutation in $O(\frac{d}{g}\log g)$ slots.
\end{corollary}
\begin{proof}
To get the claim, it is enough to plug the sorting algorithm of Theorem~\ref{thm:deterministico}
into Stage~1 of the deterministic routing algorithm proposed in~\cite{ds-IEEETPDS03}.
\end{proof}

This algorithm is not very practical. Indeed, it is based on the AKS network
that, in spite of being optimal, is not efficient when $n$ is small due to very large hidden
constants. However, the result is important from a theoretical point of view because of two facts:
it establishes that,
in principle, $O(\frac{d}{g}\log g)$ slots are enough to solve deterministically the online permutation
routing problem; and, when $d=O(g)$ and under proper hypothesis, it matches one of the lower
bounds for deterministic algorithms in the next section.

\subsection{A Few Lower Bounds}

Borodin et al.~\cite{brsu-JACM97} study the extent to which both complex hardware and
randomization can speed up routing in interconnection networks.
One of the questions they address is how \emph{oblivious routing algorithms}
(in which the possible paths followed by a packet depend only on its own source and destination)
compare with \emph{adaptive routing algorithms}. Since oblivious routing 
can usually be implemented by using limited hardware resources on each node,
it is important to understand whether
it is worth using the more complex hardware required by adaptive routing. Here, we address similar
questions. In the following, our discussion will be limited to the case $d=\Theta(g)$. 
  
Unfortunately, the concept of oblivious routing does not seem
to be useful for POPS networks.
Indeed, by adapting the ideas first used in~\cite{bh-JCSS85},
we can prove that any oblivious deterministic routing algorithm
needs $\Omega(\sqrt{g})$ slots to deliver correctly every permutation.
Moreover, by customizing and slightly adapting the approach
developed in~\cite{brsu-JACM97} (that makes use of  Yao's minimax principle~\cite{y-FOCS77}),
it is also possible
to show that any oblivious randomized routing algorithm must use $\Omega(\log g/\log\log g)$ slots on
the average.
\begin{theorem}   \label{the:obliviousDet}
For any $\POPS$ network, $d=\Theta(g)$, and any oblivious deterministic routing algorithm,
there is a permutation for which the routing time is $\Omega(\sqrt{g})$ slots. 
\end{theorem}
\begin{proof}
   We essentially customize the proof in~\cite{bh-JCSS85}
   to POPS networks,
   but also some minor modifications are in order
   to allow for passive devices and a few different assumptions.

   We assume $d=g$,
   the extension to $d=\Theta(g)$ or wider
   involving no further ideas, only more technical fuss.
   Consider the bipartite digraph $D=(V,A)$
   having the set $P$ of processors
   and the set $C$ of couplers as color classes
   and having as arcs in $A$ those pairs $(p,c)$
   such that processor $p$ can send to coupler $c$
   plus those pairs $(c,p)$
   such that processor $p$ can listen from coupler $c$.
   We have $|P|=n=dg=g^2$ processors and $|C|=g^2=n$ couplers,
   $|V|=|P|+|C| = 2n$;
   all nodes have in-degree and out-degree both equal to $g$.

   Every oblivious algorithm defines a directed $a,b$-path,
   denoted with $(a,b]$, for every pair $(a,b)\in P^2$,
   namely, the directed path of $D$
   followed by a packet with destination in $b$
   and origin in $a$.
   The characteristic vector $\chi_{(a,b]}$
   of a path $(a,b]$ is defined
   by regarding the path has the set of its nodes
   including $b$ but not $a$.   
   The {\em congestion} of a family $\Pi$ of directed paths
   is defined as $c(\Pi):=\max_{v\in V} \sum_{(a,b] \in \Pi} \chi_{(a,b]}(v)$.
   It is clear that the congestion of $\Pi$ gives a lower bound
   on the number of steps required to move a packet
   along each path in $\Pi$ since no processor in $P$ and no coupler in $C$
   can receive more than one different packet within a single slot.
   To prove the theorem we do the following:
   with reference to the path family $\{(a,b] \, | (a,b)\in P^2\}$
   determined by the oblivious algorithm under consideration,
   we show how to construct a permutation
   $\pi:P\mapsto P$ such that
   $c(\{(a,\pi(a)] \; | a\in P\}) \geq \sqrt{g}/2$.
   This will imply the stated lower bound regardless
   of the queueing discipline,
   however omniscent, employed by the algorithm.
   For every $b\in P$,
   let
   $S_b := \{v\in V\; |
           \sum_{a\in P\setminus \{b\}} \chi_{(a,b]}(v) \geq \sqrt{g}/2 \}$.
   Clearly, every path $(a,b]$, $a\notin S_b$,
   must have a last node not in $S_b$.
   Moreover, since $b\in S_b$,
   the next node on the path $(a,b]$ must be in $S_b$.
   Let $X_b$ be the set of these last nodes
   when $a$ ranges in $P\setminus S_b$.
   By definition of $S_b$,
   no node in $X_b$ can be the last node outside $S_b$
   for more than $\sqrt{g}/2$ such paths,
   hence $|P\setminus S_b| \leq |X_b|(\sqrt{g}/2)$,
   which implies $|S_b|\geq \sqrt{g}$
   in case $|X_b| < g\sqrt{g}$.
   Moreover, $|X_b| \leq g|S_b|$ since the in-degree of the network
   is bounded by $g$.
   This implies $|S_b|\geq \sqrt{g}$
   in the complementary case that $|X_b| \geq g\sqrt{g}$.
   In conclusion, $|S_b|\geq \sqrt{g}$ holds for every $b\in P$.
   Therefore, by an averaging argument,
   there must exist a $v\in V$
   which belongs to at least $\frac{|P| \sqrt{g}}{|V|}=\frac{\sqrt{g}}{2}$
   of these sets $S_b$, $b\in P$.
   Let $B=\{b\in P \,| v\in S_b\}$.
   Let $b_1, b_2, \ldots, b_{\sqrt{g}/2}$ be distinct processors in $B$
   and run the following greedy algorithm where for all processors
   $p$ in $P$ the value $\pi(p)$ is initially undefined.

\begin{quote}
   For $i:=1$ to ${\sqrt{g}/2}$,
   let $a$ be any processor in $S_{b_i}$
   such that $\pi(a)$ is undefined and define $\pi(a):=b_i$.
\end{quote}

   Notice that such an $a$ can be found at each step $i\leq {\sqrt{g}/2}$
   since at step $i$ at most $i$ values of $\pi$ have been defined,
   while $S_{b_i} \geq \sqrt{g}$.
   Moreover, $\pi$ can be clearly extended to a full permutation,
   while already
   $c(\{(a,\pi(a)] \; | \mbox{$\pi(a)$ is defined}\}) \geq
   |\{a\, | \mbox{$\pi(a)$ is defined}\}| = \sqrt{g}/2$
   since node $v$ belongs to each path $(a,\pi(a)]$ by construction.
\end{proof}

\begin{theorem}   \label{the:averageInput}
For any $\POPS$ network, $d=\Theta(g)$, and any oblivious deterministic routing algorithm,
the expected routing time for a random permutation (with each permutation chosen with uniform probability) is $\Omega(\log g/\log\log g)$. 
\end{theorem}
\begin{proof}
   The proofs to be customized and adapted here come
   from~\cite{brsu-JACM97}.
   The customization starts again by considering
   the bipartite digraph $D=(V,A)$
   on color classes $P$ and $C$
   introduced in the proof of Theorem~\ref{the:obliviousDet}.
   Also the various small adjustment
   are in analogy with those detailed
   in the proof of Theorem~\ref{the:obliviousDet}.
\end{proof}

\begin{corollary}
For any $\POPS$ network, $d=\Theta(g)$ and any oblivious deterministic routing algorithm,
there is a permutation for which the expected routing time is $\Omega(\log g/\log\log g)$. 
\end{corollary}
\begin{proof}
   To get this corollary of Theorem~\ref{the:averageInput},
   use the Yao's minimax principle~\cite{y-FOCS77}
   in perfect analogy to what is done in~\cite{brsu-JACM97}.
\end{proof}

These complexities are not satisfactory. Indeed, here in this paper we show a non-oblivious
deterministic algorithm
that runs in $O(\log g)$ slots and a non-oblivious randomized one that runs in $O(\log\log g)$ slots
with high probability.
So, by restricting to oblivious algorithms, it may be true that we get a (somewhat) simpler processor,
but we also lose
an exponential factor in running time, both with and without randomization. This is not a
good deal.
Therefore, we will not discuss oblivious routing any more, and will focus only on
adaptive routing.

Finding good lower bounds for adaptive deterministic routing is not trivial.
In~\cite{brsu-JACM97}, the authors explicitly say that they were not able to
provide any result for this case in their context. Here, we give partial answers.
First, we prove a $\Omega(\log g)$ tight lower bound for a special case of adaptive deterministic routing
that applies both to the hypercube simulation routing algorithm in~\cite{ds-IEEETPDS03}
and to our deterministic algorithm (that is, in this context, optimal). Second, we prove
a strong separation theorem between determinism and randomization. Indeed, we can show
both a $\Omega(\log g)$ lower bound for a class of adaptive deterministic routing algorithms,
and a $O(\log\log g)$ upper bound for the same class where processors are allowed to
generate and use randomization.
To the best of our knowledge, this is the first separation theorem showing a gap between
$\log n$ and $\log\log n$.

Consider our deterministic routing algorithm, proposed in the previous section. It is based
on a simulation of the AKS sorting network. At every slot, each processor sends its packet
to a pre-determined other processor, according to
the comparator it is going to simulate in the slot. So, the communication patterns are fixed for
the whole computation, and do not depend on the input permutation. We can prove
a lower bound for all algorithms that have the same property. More formally,
a routing algorithm is called \emph{rigid} if, at every slot~$t$, each processor~$i$ sends one of the
packets it currently stores to the set of groups~$C_{\mathrm{out}}(i,t)$, and listens to
group~$c_{\mathrm{in}}(i,t)$, where functions $C_{\mathrm{out}}$ and $c_{\mathrm{in}}$ depend
solely on $t$ and on the processor index.
Here, we can assume that the processors have enough local memory to store a copy of all the
packets they have seen so far and that
they choose the packet to send according to any strategy or algorithm.
This is enough to get the following lower bound.
\begin{theorem}
Any deterministic and rigid algorithm for online permutation routing
on the ${\mathrm{POPS}}(d,g)$ network, $d=\Theta(g)$, must use $\Omega(\log n)$ slots.
\end{theorem}
\begin{proof}
Consider a processor~$i$. Let $P(i,t)$ be the set of all packets that are potentially stored
by processor~$i$ at slot~$t$, according to the routing algorithm.
At the beginning, $P(i,0)=\{p_i\}$. During slot~$t$, processor~$i$
can receive at most one packet from group~$c_{\mathrm{in}}(i,t)$. Assume this packet
comes from processor~$j$. Index~$j$ is statically determined and is independent
of the initial permutation, since the algorithm is rigid.
So, either $P(i,t)=P(i,t-1)\cup P(j,t-1)$ or $P(i,t)=P(i,t-1)$, if no packet is sent to
group~$c_{\mathrm{in}}(i,t)$ (because there is no such processor~$j$, or a conflict
occurred). Therefore, $|P(i,t)|\le 2^t$ for all $t\ge 0$.

Now, assume that the algorithm stops after $t<\log n$
slots. Then, $|P(i,t)|<n$, and there exists $h$ such that $p_h\notin  P(i,t)$. As a consequence,
the routing algorithm must fail for all input permutations such that the destination of
$p_h$ is processor~$i$. We conclude that $t=\Omega(\log n)$.
\end{proof}
This bound applies to both the $O(\log^2 g)$ algorithm in~\cite{ds-IEEETPDS03}
and to our deterministic algorithm in the previous section.
Therefore, within the class of rigid algorithms, our proposed routing scheme is optimal.

Now, we prove a strong separation theorem. Under restricted hypotheses, we can show that
randomization can give an exponential speed-up over determinism. Here, we address a class
of routing algorithms we call \emph{two-hops algorithms}. A two-hops algorithm has the
following properties: 
\begin{enumerate}
\item
Every processor has two buffers, an $A$-buffer and a $B$-buffer;
\item
at the beginning, the packets are stored in the $A$-buffer of each processor;
\item
at every odd slot~$2t+1$, $t=0,1,\dotsc$, every processor~$i$ with a packet in the $A$-buffer
sends the packet to group~$c_{\textrm{out}}(i,2t+1)$ (two-hops algorithms can only use unicast),
listens to incoming packets from
group~$c_{\textrm{in}}(i,2t+1)$, and store the incoming packet (if any) into the $B$-buffer;
\item
at every even slot~$2t$, $t=1,\dotsc$, every processor~$i$ sends the packet in the $B$-buffer to
destination, reset the $B$-buffer, and listens to incoming packets from coupler~$c_{\textrm{in}}(i,2t)$.
\end{enumerate}
Also, we will make the following assumptions:
\begin{enumerate}
\addtocounter{enumi}{4}
\item
when multiple packets use the same coupler (multiple packets from a group sent to the
same group), no packet is delivered.
\item
When a packet arrives to any processor in the destination group, it is considered to be
successfully routed, and disappears from the network (from the original $A$-buffer as well);
\end{enumerate}
The last hypothesis simplifies the job of routing all the packets to destination---we don't
have to take care of acks when packets reach their destination. However,
since we are proving a lower bound, we don't lose generality. Now, our goal is to show that
for every deterministic choice of functions $c_{\textrm{in}}$ and $c_{\textrm{out}}$, there exists
an input permutation such that the routing is completed in $\Omega(\log g)$ slots. On the other
hand, our randomized algorithm shows that there exists a deterministic $c_{\textrm{in}}$ and
a randomized $c_{\textrm{out}}$ such that all the packets are routed to destination in
$O(\log\log g)$ slots with high probability.

Consider a deterministic two-hops algorithm. Assume that the algorithm stops
after $T<\frac{1}{2}\min\{\log d, \log g\}$ slots, $T$ even. We will say that processor~$i$
\emph{shoots} on group~$a$ in the first $T$ slots if there exists an odd $t<T$
such that $c_{\textrm{out}}(i,t)=a$.
\begin{lemma}
There exists a group $a_0$ such that at most $dT$ processors shoot on $a_0$
in the first $T$ slots.
\end{lemma}
\begin{proof}
By counting.
\end{proof}
\begin{corollary}
\label{cor:separation}
There are at least $n-dT=dg-dT>dg/2$ processors~$i$ such that
processor~$i$ does not shoot on $a_0$ in the first $T$ slots.
\end{corollary}
Let $P(a_0)$ be the set of processors~$i$ such that processor~$i$ does not shoot
on $a_0$ in the first $T$ slots. By Corollary~\ref{cor:separation}, $|P(a_0)|>dg/2$.
A subset $A\subset P(a_0)$ is \emph{$\sqrt{g}$-robust} if for every $i\in A$ and
for every $t<T$ there are at least $\sqrt{g}$ processors~$j$ in $A$ such that
$c_{\textrm{out}}(i,t)= c_{\textrm{out}}(j,t)$.
\begin{lemma}
There exists a $\sqrt{g}$-robust subset $P'(a_0)\subset P(a_0)$ such that
$|P'(a_0)|\ge \frac{dg}{2}-Tg\sqrt{g}$.
\end{lemma}
\begin{proof}
If $P(a_0)$ is not $\sqrt{g}$-robust, then there must be a processor~$i\in P(a_0)$
and a $t<T$ such that $c(i,t)=c(j,t)$ for less than $\sqrt{g}$
processors~$j\in P(a_0)$. This means that all the processors~$j$ such that $c(i,t)=c(j,t)$
(including $i$) must be removed from $P(a_0)$ to get a $\sqrt{g}$-robust subset. So,
let $P_1(a_0)$ be obtained from $P(a_0)$ by removing all these processors and mark
the pair $(t,c(i,t))$. Start now from $P_1(a_0)$ in place of $P(a_0)$ and keep iterating.
Notice that no pair can be marked twice in the process. The number of pairs is at most
$Tg$, and each time we mark a pair we drop at most $\sqrt{g}$ processors.
\end{proof}
\begin{theorem}
Any deterministic and two-hops algorithm for online permutation routing
on the ${\mathrm{POPS}}(d,g)$ network, $d=\Theta(g)$, must use $\Omega(\log n)$ slots.
\end{theorem}
\begin{proof}
We will show that for every processor~$i$ in $P'(a_0)$ there exists an input permutation
such that $p_i$ will not reach destination. The idea of the proof is as follows: we can build
an input permutation such that $p_i$ has to perform two hops to get to destination,
and that has a conflict at every even slot. Take a packet~$p_i$ such that $i\in P'(a_0)$
and mark the packet. Now, for $t:=T-1$ downto $1$, $t$ odd, do the following:
\begin{quote}
for every marked packet~$p_j$,
\begin{enumerate}
\item
take an unmarked packet~$p_h$ such that $c(h,t)=c(j,t)$;
\item
mark packet~$p_h$.
\end{enumerate}
\end{quote}
Then, set the destination of all marked packets to processors in group~$a_0$, so that
no marked packet can get to destination in one hop (they are chosen from $P'(a_0) \subseteq P(a_0)$).
The number
of packets that are marked in the above process does not exceed $d$ nor $\sqrt{g}$,
since $T<\frac{1}{2}\min\{\log d, \log g\}$. The important property guaranteed by the above
process is that any packet~$p_j$ marked at time~$t$ will experience a conflict during all
even slots from the beginning of the routing to time~$t$. In particular, packet~$p_i$ does
not reach destination within $T=\Omega(\log n)$ slots.
\end{proof}

We believe that the $\Omega(\log g)$ lower bound for deterministic routing holds in
a much wider setting. This is described in the following two conjectures.
\begin{conjecture}
There exists a deterministic algorithm for online permutation routing
on the ${\mathrm{POPS}}(d,g)$ network, $d=\Theta(g)$, that is optimal and conflict-free.
\end{conjecture}
\begin{conjecture}
Any deterministic and conflict-free algorithm for online permutation routing
on the ${\mathrm{POPS}}(d,g)$ network, $d=\Theta(g)$, must use $\Omega(\log n)$ slots.
\end{conjecture}

\section{A Randomized Algorithm}

Here we present our randomized algorithm. In the following, we will make use
of the so called \emph{union bound}, a simple bound on the union of events.
\begin{fact}[Union Bound]
Let $E_1,\dotsc,E_m$ be $m$ events. Then,
\begin{equation*}
\Pr\left[\bigcup_{i=1}^m E_i\right]\le \sum_{i=1}^m \Pr\left[E_i\right].
\end{equation*}
\end{fact}
We will use a function $\Delta(x):=x \mod g$.
Moreover, we will say that some event happens \emph{with high probability}
meaning that the probability of the event is $1-1/g^k$ for some positive $k$.

\subsection{The Case $d=g$}

Given a packet $p_i$, $i\in{\mathbb{N}}_{n}$, its \emph{temporary destination group} is
group~$\Delta(\pi(i))=\pi(i)\mod g$.
Note that there are exactly $d$ packets with temporary destination group~$a$,
for all $a\in{\mathbb{N}}_{g}$.
The idea of the routing algorithm is as follows: Each packet is first routed to
a randomly and independently chosen \emph{random intermediate group}, then to its
temporary destination group, and lastly to its final destination.
So, we iterate the following \emph{step}, composed of five slots:
\begin{figure*}
\centering\includegraphics[scale=.7]{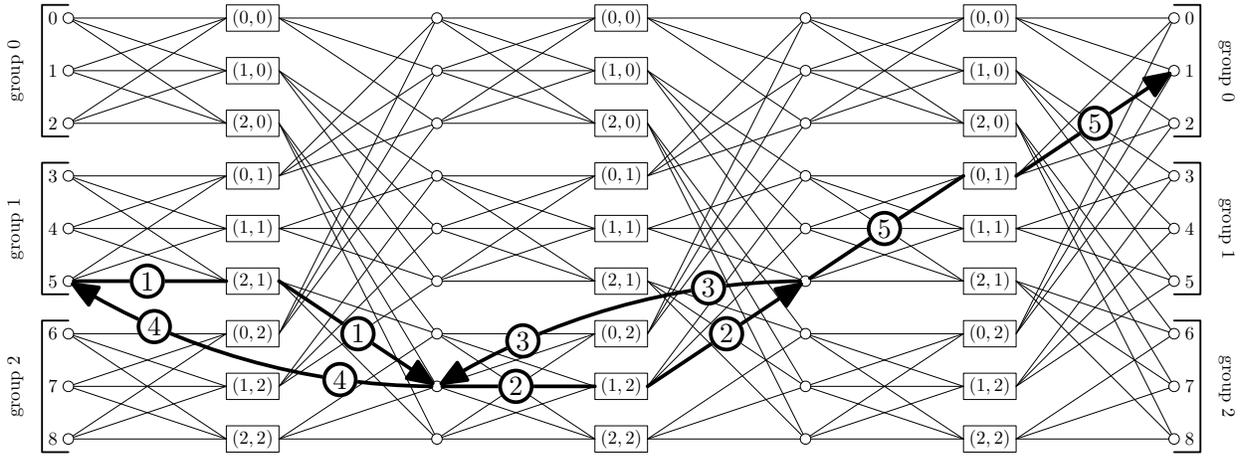}
\caption{Example of randomized routing in a ${\mathrm{POPS}}(3,3)$ network. Packet~$p_5$ has destination $\pi(5)=1$ in group~$0$.
Its temporary destination group is group $\pi(5)\mod g=1$. In this step, the
random intermediate group chosen by packet~$p_5$ is group~$2$.}
\end{figure*}
\begin{enumerate}
\item
each processor containing a packet~$p$ to be routed chooses a random intermediate group~$r$
(uniformly and independently at random over ${\mathbb{N}}_{g}$)
and sends a copy of packet~$p$ to group~$r$;
\item
every copy that arrived to the random intermediate group is sent to its temporary destination group;
\item
for each copy that arrived to the temporary destination group an ack is sent back
to the random intermediate group;
\item
for each ack arrived to the random intermediate group, an ack is sent back to the source processor which, in turn, deletes the original packet;
\item
every copy that arrived to its temporary destination group is sent to its destination.
\end{enumerate}
During the step, there are at most two replicas of the same packet. One is the \emph{original
packet}, stored in the source processor; the other is the \emph{copy}, that tries to go from
the source processor to a random intermediate group, then to its temporary destination
group, and finally to its destination. In slot~4, if the source processor receives an ack, it
can be sure that the copy has been successfully delivered, as proved in
Proposition~\ref{pro:conflictless}, and can safely delete the original packet.
In fact, the original packet gets deleted in slot~4 if and only if, within the step, the copy
gets to destination in slot~5.

In slots~1, 2, and~5, for every group~$a$, every processor~$i$ in group~$a$ is responsible for
listening to coupler~$c(a, \Delta(i))$
for the message possibly coming from
group~$\Delta(i)$.
This way, every
conflict-less communication successfully completes and no packet is lost. Indeed,
during slots~1 and~2, in every group $a$, $a\in{\mathbb{N}}_{g}$, the processor with index~$b$
within the group, $b\in{\mathbb{N}}_{g}$, receives the packet that is possibly coming
from group~$b$. In slot~5, every processor~$\pi(i)$ that still has to receive packet~$p_{i}$
hopefully receives its packet from group~$\Delta(\pi(i))$, the temporary destination group of
packet~$p_i$.
Slots~3 and~4 behave differently. Indeed, each ack sent during slot~3 is received by the
same processor that sent the packet in slot~2. Similarly, each ack sent during slot~4 is received by the same processor that sent the packet in slot~1.

Clearly, during slots~1 and~2, multiple conflicts on the couplers should be expected,
and many of the communications may not complete. For example, two packets in the same
group can choose the same random intermediate group during slot~1, or two packets willing
to go to the same temporary destination group are currently in the same random intermediate
group during slot~2.
On the contrary, slots 3, 4, and 5 do not generate any conflict,
as shown in the following proposition.
\begin{proposition}
\label{pro:conflictless}
At all steps, slots 3, 4, and 5 of the routing algorithm do not generate any conflict.
\end{proposition}
\begin{proof}
Consider packet~$p_i$ stored at processor~$i$ in group~$a$. Assume that, during an arbitrary
step, its random intermediate group is $r(i)$, chosen
uniformly
at random.
In the case when packet~$p_i$ survives slot~1 and arrives to its random intermediate group
$r(i)$, we know that coupler $c(r(i),a)$ has been used to send
packet~$p_i$ only, otherwise a conflict would have stopped the packet.
Moreover, since there is only one processor in group~$r(i)$ that is responsible
for receiving packet~$p_i$, namely processor~$r(i)d+a$,
there will be only one ack message corresponding to packet~$p_i$ to be sent in slot~4,
and this ack message is the only one that uses the symmetric coupler
$c(a,r(i))$ during slot~4.
In conclusion, slot~4 is conflict-free.
A similar argument shows that slot~3 is conflict-free as well.

Consider now slot~5. Assume that, after step~4, packet $p_j$ has arrived at the 
same temporary destination group as packet~$p_i$.
This means that $\Delta(\pi(i))=\Delta(\pi(j))$. That is,
$\pi(i)\equiv \pi(j)\mod g$. In this case, it is not possible that $\pi(i)$ and $\pi(j)$ are
in the same group; otherwise we would have 
$\pi(i)=\pi(j)$, in contrast with the fact that  $\pi$ is a permutation.
Therefore, packets~$p_i$ and $p_j$ go to different groups from their temporary
destination group. In other words, step~5 is conflict-free as well. 
\end{proof}

By Proposition~\ref{pro:conflictless}, if packet~$p_i$ survives the first two slots of a step,
then, in the very same step, it will be routed to its destination, and an ack will be successfully
returned to source processor~$i$. When the ack arrives, the source processor can delete
the packet, since it knows it will be safely stored by the destination processor.
Conversely, if no ack arrives, the packet is not deleted, and the processor
tries again to deliver it in the next step, choosing again a possibly different
random temporary group.

\begin{figure*}
\centering
\subfigure[Conflict graph~$G_{\pi}$;]{\label{fig:bipartite-1}
\includegraphics[scale=.78]{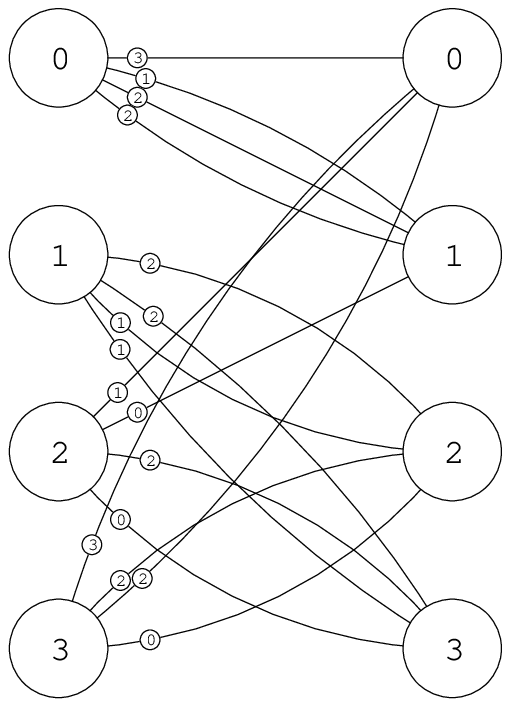}}\goodgap
\subfigure[conflict graph~$G_{\pi}$, where only packets surviving slot~1 are shown;]{\label{fig:bipartite-2}
\includegraphics[scale=.78]{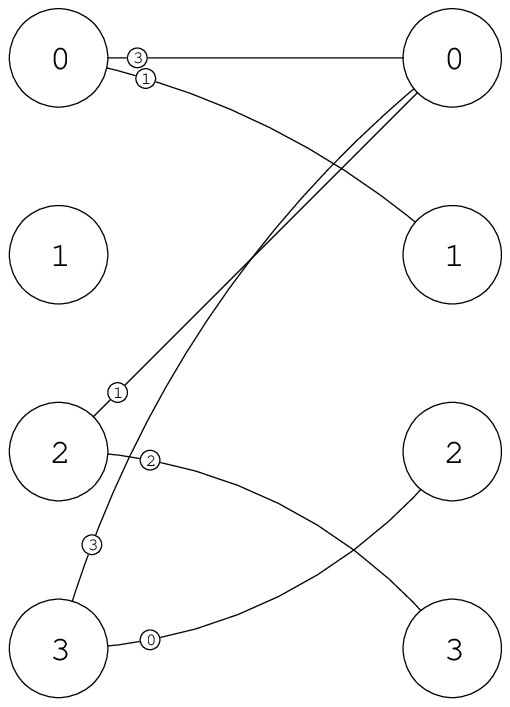}}\goodgap
\subfigure[conflict graph~$G_{\pi}$, where only packets surviving both slot~1 and slot~2 are shown.]{\label{fig:bipartite-3}
\includegraphics[scale=.78]{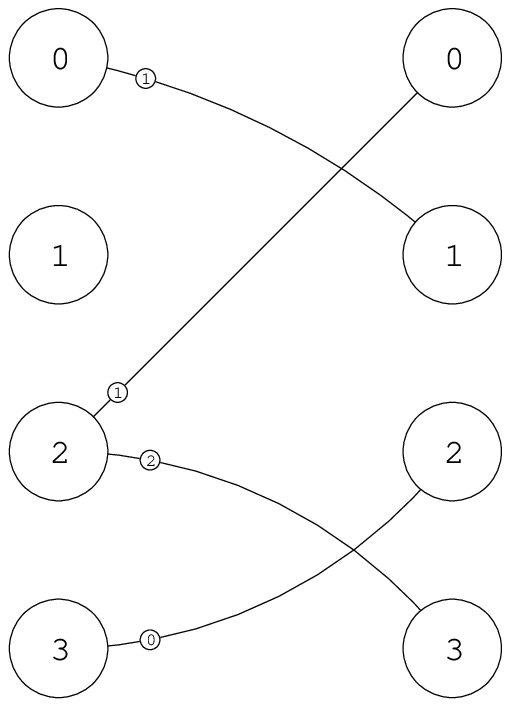}}
\caption{Conflict graph~$G_{\pi}$, where permutation~$\pi=[1,5,8,9,3,10,11,14,15,13,0,7,2,6,12,4]$
(consequently, $\Delta(\pi(\cdot))=[1,1,0,1,3,2,3,2,3,1,0,3,2,2,0,0]$), in a ${\mathrm{POPS}}(4,4)$
network.}
\label{fig:bipartite}
\end{figure*}

By the above discussion, we can safely concentrate on slots~1 and~2. 
A useful way to visualize the conflicts in slots~1 and~2
of an arbitrary step is
shown in Figure~\ref{fig:bipartite-1}.
At any given step of the routing algorithm, let $\pi$ be the
restriction of the input permutation to those packets that have not been successfully routed yet (during previous steps).
We build the \emph{graph of conflicts}, a bipartite multi-graph $G_{\pi}$ on node classes
$S:={\mathbb{N}}_g$ and $D:={\mathbb{N}}_g$. For every group~$a$ and for each
packet~$p_i$ in group~$a$ and yet to be
routed, we introduce an edge with one endpoint in $a\in S$ and the other endpoint in
the temporary destination group~$\Delta(\pi(i))\in D$.
During slot~1 of the step,
every edge (packet yet to be routed) randomly and uniformly chooses a
\emph{color} in ${\mathbb{N}}_g$ (the random intermediate
group).
Clearly, a same packet can choose different colors
in different steps of the routing algorithm.
Now we can exactly characterize the conflicts in the first two slots of the routing algorithm during
step~$s$.
Packet~$p_i$ in group~$a$ (represented by an edge from $a\in S$ to $\Delta(\pi(i))\in D$)
has a conflict during slot~1
if and only if there is another edge incident to $a\in S$ with
the same random color. Moreover, if we remove all edges relative to packets that have a conflict
in slot~1 (see Figure~\ref{fig:bipartite-2}), every remaining packet $p_i$ has a
conflict during slot~2
if and only if there is another remaining edge incident to $\Delta(\pi(i))\in D$ with the same random color.
Figure~\ref{fig:bipartite-3} shows which packets of Figure~\ref{fig:bipartite-1} survive both
slots and are hence delivered to destination by Proposition~\ref{pro:conflictless}. 

Our first result shows that, in case the packets are ``sparse'' in the network,
then all the packets can be delivered in a constant number of slots with high
probability.
\begin{lemma}
\label{lem:phase3}
If the maximum degree of the conflict graph is $g^{\alpha}$
for some constant $\alpha<1$, then the routing algorithm delivers all the packets to destination in a
constant number of slots with high probability.
\end{lemma}
\begin{proof}
Since the maximum degree of the conflict graph is $g^{\alpha}$, in every group of the POPS network
there are at most $g^{\alpha}$ packets left to be routed, and every group of the POPS network is
the temporary destination group of at most $g^{\alpha}$ packets.
Let $\beta=1-\alpha$.
We show that
the probability 
that all packets get routed to destination
within $3/\beta$ steps is at least $1-c_\beta/g$,
where $c_\beta := 2^{3/\beta}$ is a constant
depending only on (the constant) $\beta$.
Consider a generic packet $p_i$ in group~$a$.
The probability that packet~$p_i$ has a conflict in one step is at most equal to the
probability that either one of the packets in group~$a$ or one of the packets
with temporary destination group~$\Delta(\pi(i))$ chooses the same random intermediate group as
packet~$p_i$.
Since at most $g^\alpha-1$ other packets are in group~$a$, and similarly at
most $g^\alpha-1$ have temporary destination group~$\Delta(\pi(i))$, this probability cannot be larger
than $2g^\alpha/g=2g^{-\beta}$.
Therefore, the probability that the packet is not routed
in each of the $3/\beta$ steps is at most
\begin{equation*}
\left(\frac{2}{g^\beta}\right)^{\frac{3}{\beta}}=\frac{2^{3/\beta}}{g^3}=\frac{c_\beta}{g^3}.
\end{equation*}
By the union bound, the probability that any of the $g^{1+\alpha}<g^2$ packets in the network
has not been routed in $3/\beta$ steps is at most $c_\beta/g$.
\end{proof}
As a matter of fact, the hard part of the job is to reduce the initial number of $g$ packets in each
group in such a way to get a ``sparse'' set of remaining packets.
We can prove that this is done quickly
by our randomized algorithm by
providing sharp bounds on the number $X$ of packets that are
successfully delivered in a step.
We define $X$ as a sum of indicator random variables $Z_i$, where
$Z_i$ is equal to $1$ if the $i$-th packet is delivered in this step, and $0$ otherwise.
It is important
to realize that these random variables are not independent: the event that one packet has
a conflict influences the probability that another packet has a conflict as well.
As a consequence, we cannot use the well-known Chernoff bound
to get sharp estimates of the value of $X$ since there does not seem to be any
way to describe the process as a sum of independent random variable.
So, we need a more sophisticated 
mathematical tool.
Specifically, we will see that slots~1 and~2 of one step
of the routing algorithm can be modeled by a set of martingales. Martingale theory is 
useful to get sharp bounds when the process is described in terms of not necessarily
independent random variables.

For an introduction to martingales, the reader is 
referred to~\cite{mr95}.
Also~\cite{ds65}, \cite{gs88}, \cite{as00}, and~\cite{dp04} give a description of martingale
theory. Here, we give a brief review of the main definitions and theorems
we will be using in the following.
\begin{definition}[\cite{mr95}]
Given the $\sigma$-field $(\Omega, {\mathbb{F}})$ with ${\mathbb{F}}=2^{\Omega}$, a
\emph{filter} is a nested sequence
${\mathbb{F}}_0\subseteq {\mathbb{F}}_1\subseteq \dotsb \subseteq {\mathbb{F}}_m$ of
subsets of $2^{\Omega}$ such that
\begin{enumerate}
\item
${\mathbb{F}}_0=\{\emptyset, \Omega\}$;
\item
${\mathbb{F}}_m=2^{\Omega}$;
\item
for $0\le h\le m$, $(\Omega, {\mathbb{F}}_h)$ is a $\sigma$-field.
\end{enumerate}
\end{definition}
\begin{definition}[\cite{mr95}]
Let $(\Omega, {\mathbb{F}}, \PR)$ be a probability space with a filter
${\mathbb{F}}_0,\dotsc, {\mathbb{F}}_m$. Suppose that $Y_0, \dotsc, Y_m$ are random variables
such that for all $h\ge 0$, $Z_h$ is ${\mathbb{F}}_i$-measurable. The sequence $Z_0,\dots,Z_m$
is a \emph{martingale} provided that, for all $h\ge 0$,
\begin{equation*}
\E[Z_{h+1}|{\mathbb{F}}_h]=Z_h.
\end{equation*}
\end{definition}
The next tail bound for martingales is similar to the Chernoff bound for the sum of Poisson
trials.
\begin{theorem}[Azuma's Inequality~\cite{mr95}]
Let $Z_0,\dotsc,Z_m$ be a martingale such that for each $h$,
\begin{equation*}
|Z_h-Z_{h-1}|\le c_h,
\end{equation*}
where $c_h$ may depend on $h$. Then, for all $t\ge 0$ and any $\lambda>0$,
\begin{equation*}
\PR\left[ |Z_t-Z_0|\ge\lambda\right]\le 2e^{-\frac{\lambda^2}{2\sum_{k=1}^t c_k^2}}.
\end{equation*}
\end{theorem}
\begin{theorem}
\label{thm:fondamentale}
A $\POPSg$ network can route any permutation in $O(\log\log g)$ slots
with high probability.
\end{theorem}
\begin{proof}
Let $G_{\pi}=(S,D; E)$ be the conflict graph at step~$s$ of the routing
algorithm, where $\pi$ is the input permutation restricted to those packets
that still have to be routed at the beginning of step~$s$.
Let $d_s$ be the maximum degree of $G_{\pi}$.
So, at step~$s$ there are at most $d_s$ packets left to
be routed in every group, and at most $d_s$ packets are willing to go to
the same temporary destination group.
Clearly, $d_1\leq d$. We will show that after
$O(\log\log g)$ steps the conflict graph has maximum degree at most $g^{5/6}$.
This is enough to prove this theorem by Lemma~\ref{lem:phase3}.

Assume to be at step~$s$. If $d_s\le g^{5/6}$, then we are done.
So, we can assume that $d_s> g^{5/6}$.
Let $S_a$, $a\in S$, be the set of indices
of the packets of group~$a$ that still have to be delivered at the beginning of
step~$s$. Similarly, let $D_b$, $b\in D$,
be the set of indices
of the packets in the whole network that still have to be delivered and that have group~$b$ as
temporary destination group.
Clearly, $|S_a|$ and $|D_b|$ are the degrees of nodes $a\in S$ and $b\in D$ in the conflict graph of step~$s$.
Therefore, $|S_a|\le d_s$ and $|D_b|\le d_s$ for every $a\in S$ and $b\in D$.
For every packet~$p_i$ still to be routed, we define the following indicator random variable,
\begin{equation*}
Z_i^1=\begin{cases} 1 & \textrm{if packet~$p_i$ survives slot~1 in step~$s$,}\\
0 & \textrm{otherwise.} \end{cases}
\end{equation*}
Random
variable~$X_a^1=\sum_{i\in S_a} Z_i^1$ tells the number of packets from group~$a$ that
survive slot~1; random variable~$Y_b^1=\sum_{j\in D_b} Z_j^1$ tells the number of packets with temporary destination
group~$b$ that survive slot~1.
Moreover, let random
variable~$C_i$ be equal to the color chosen by packet~$p_i$ in step~$s$.

Clearly, we have nothing to show about the nodes in $G_{\pi}$ that have degree smaller
than or equal to $g^{5/6}$. So, we define sets $S^+\subseteq S$ and
$D^+\subseteq D$, which collect the nodes with degree
larger that $g^{5/6}$, and focus on the nodes in these sets.
Consider an arbitrary node $a\in S^+$.
The expectation of $Z_i^1$, $i\in S_a$, can be bounded as follows:
\begin{equation}
\begin{split}
\E[Z_i^1] = \PR[\forall \; h\in S_a\setminus \{i\}, \; C_h\neq C_i]=
\prod_{h\in S_a\setminus \{i\}} \PR[C_h\neq C_i]\\
= \left( 1-\frac{1}{g}\right)^{|S_a|-1}
\ge e^{-|S_a|/g}.
\end{split}
\label{eqn:lowerEXi}
\end{equation}
So, the expected number of packets in group~$a$ that survive slot~1 can be bounded accordingly,
\begin{equation}
\label{eqn:lowerX}
\E[X_a^1]=\E\left[\sum_{i\in S_a} Z_i^1\right]=\sum_{i\in S_a} \E[Z_i^1]\ge |S_a| e^{-|S_a|/g}.
\end{equation}

In order to show that random variable $X_a^1$ is not far from its expectation with high probability,
we now define random variables $W_h=\E[X_a^1|{\mathbb{F}}_h]$, $h=0,\dotsc,|S_a|$,
where ${\mathbb{F}}_h$ is the $\sigma$-field generated by the random color chosen by
the first $h$ packets in $S_a$.
Filter ${\mathbb{F}}_h$, $h=0,\dotsc,|S_a|$, is such that $W_0, \dotsc, W_{|S_a|}$ is a martingale and that
$|W_{h}-W_{h-1}|\le 2$,
since fixing the random color chosen by the $h$-th packet in $S_a$
can only affect the expected value of the sum $X_a^1$ at most by two.
By the Azuma's inequality, for every $\delta>0$
\begin{equation}
\begin{split}
\PR\left[\left|X_a^1-\E[X_a^1]\right|\ge \delta\E[X_a^1]\right]=\PR\left[\left|W_{|S_a|}-W_0\right|
\ge \delta\E[X_a^1]\right]\\
\le 2e^{-\frac{\delta^2 \E[X_a^1]^2}{2\sum(2)^2}}
\le 2e^{-\frac{\delta^2 |S_a|^2e^{-2d_s/g}}{8|S_a|}}\le 
2e^{-\frac{\delta^2 g^{5/6}}{8e^2}}.
\end{split}
\label{eqn:azumaX}
\end{equation}

To prove a similar result for $Y_b^1$, $b\in D^+$,
we must recast the above general martingale arguments
into a more structured approach.
This is because $Y_b^1$ may depend on the random colors chosen by all the packets
in the network, and not only on those chosen by the packets in $D_b$.

Consider an arbitrary node $b\in D^+$.
In the following analysis of the expectation and concentration
of $Y_b^1$ we can clearly pretend that the random colors
are first choosen  for the packets outside $D_b$ and later for the packets in $D_b$.
This will not invalidate our conclusions about the whole of the $Y_b^1$'s, $b\in D^+$,
since these will be derived from the solid claims about any single $Y_b^1$ by the union bound.
For every $a\in S_a$, we define set~$\overline{C}_{a,\overline{b}}$
as ${\mathbb{N}}_g\setminus C_{a,\overline{b}}$, where $C_{a,\overline{b}}$ is the set of colors that are chosen in
step~$s$ by a packet in group~$a$ that has temporary destination group different from $b$,
\begin{equation*}
\overline{C}_{a,\overline{b}}={\mathbb{N}}_g\setminus\left(\bigcup_{i\in S_a\setminus D_b} \left\{ C_i \right\}\right).
\end{equation*}
The average size of $\overline{C}_{a,\overline{b}}$ is
\begin{equation*}
\E\left[\left| P_{b.a}\right|\right]=
g\left(1-\frac{1}{g}\right)^{|S_a\setminus D_b|}.
\end{equation*}
Being just a classical ball and bins problem~\cite{mr95}, we know that random variable~$|\overline{C}_{a,\overline{b}}|$
is not far from its expectation with probability
\begin{equation*}
\PR[|\overline{C}_{a,\overline{b}}|<(1-\delta)\E[|\overline{C}_{a,\overline{b}}|]\le e^{-\frac{\delta^2\E[|\overline{C}_{a,\overline{b}}|]^2}{2g}}
\le e^{-\frac{\delta^2g}{2e^2}},
\end{equation*}
for every $\delta>0$. By the union bound over the $g$ nodes in $S$, for every $\delta>0$,
we know that for every node $a\in S$
\begin{equation}
\label{eqn:lowerP}
|\overline{C}_{a,\overline{b}}| \ge (1-\delta)g\left(1-\frac{1}{g}\right)^{|S_a\setminus D_b|}
\end{equation}
with probability
\begin{equation}
\label{eqn:azumaP}
1-ge^{-\frac{\delta^2g}{2e^2}}.
\end{equation}

Under the hypothesis that Equation~\ref{eqn:lowerP} holds for every $a\in S$, we can bound the expectation of $Z_j^1$, $j\in D_b$, as follows:
\begin{equation*}
\E[Z_j^1]  = \PR\left[\left(\forall \; h\in D_b\cap S_{a_j}^{1}\setminus \{j\}, \; C_h\neq C_j\right) \wedge
(C_j\in P_{b,a_j})\right],
\end{equation*}
where $a_j$ is the group of packet~$p_j$. So,
\begin{align*}
\E[Z_j^1]  & \ge \left(1-\frac{1}{g}\right)^{|D_b\cap S_{a_j}^{1}\setminus \{j\}|}
(1-\delta) \left(1-\frac{1}{g}\right)^{|S_{a_j}^{1}\setminus D_b|}\\
& = (1-\delta)\left(1-\frac{1}{g}\right)^{|S_{a_j}^{1}\setminus \{j\}|}
\ge (1-\delta) e^{-|S_{a_j}^{1}|/g}.
\end{align*}
The expectation of $Y_b^1$ can be bounded accordingly,
\begin{equation}
\label{eqn:lowerY}
\E[Y_b^1]=\E\left[\sum_{j\in D_b} Z_j^1\right]=\sum_{j\in D_b} \E[Z_j^1]
\ge (1-\delta)|D_b| e^{-|D_b|/g}.
\end{equation}

In order to show that random variable $Y_b^1$ is not far from its expectation with high probability,
we now define random variables $W_k=\E[Y_b^1|{\mathbb{F}}_k]$, $k=0,\dotsc,|D_b|$,
where ${\mathbb{F}}_k$ is the $\sigma$-field generated by the random color
chosen by the first $k$ packets in $D_b$.
Filter ${\mathbb{F}}_k$, $k=0,\dotsc,|D_b|$, is such that $W_0, \dotsc, W_{|D_b|}$
is a martingale and that $|W_{k}-W_{k-1}|\le 2$,
since fixing the random color chosen by the $k$-th packet in $D_b$
can only affect the expected value of the sum $Y_b^1$ at most by two.
By the Azuma's inequality, for every $\delta>0$
\begin{equation}
\begin{split}
\PR\left[\left|Y_b^1-\E[Y_b^1]\right|\ge \delta\E[Y_b^1]\right]=\PR\left[\left|W_{|D_b|}-W_0\right|
\ge \delta\E[Y_b^1]\right]\le\\
\le 2e^{-\frac{\delta^2 \E[Y_b^1]^2}{2\sum(2)^2}}
\le 2e^{-\frac{\delta^2 (1-\delta)^4|D_b|^2e^{-2d_s/g}}{8|D_b|}}\le 
2e^{-\frac{\delta^2 (1-\delta)^4 g^{5/6}}{8e^2}}.
\end{split}
\label{eqn:azumaY}
\end{equation}

Let $G_{\pi'}=(S,D; E')$ be the conflict graph at step~$s$, where
$\pi'$ is the input permutation restricted to those packets that survive slot~1 in step~$s$.
Hence, $E'\subseteq E$.
Our goal is to bound the number of packets that survive slot~2
as well, and are thus delivered to destination during this step.
Let $Z_j^2$
be equal to one if packet~$p_j$ survives both slots~1 and~2,
and zero otherwise.
Also, let $S_a^1$, $a\in S$, be the set of indices
of the packets of group~$a$ that have survived slot~1. Similarly, let $D_b^1$, $b\in D$,
be the set of indices
of the packets in the whole network that have survived slot~1 and have group~$b$ as
temporary destination group. Clearly, for every $a\in S$,
$|S_a^1|$ is equal to $X_a^1$ and is the degree of node~$a$ in $G_{\pi'}$;
while for every $b\in D$, $|D_b^1|$ is equal to $Y_b^1$ and is the degree of node~$b$
in $G_{\pi'}$. 
Random variables
\begin{equation*}
X_a^2=\sum_{i\in S_a^1} Z_j^2,
\end{equation*}
$a\in S$, tell the number of packets in group~$a$ that
are delivered during step~$s$; similarly, random variables
\begin{equation*}
Y_b^2=\sum_{j\in D_b^1} Z_j^2
\end{equation*}
$b\in D$, tell the number of packets willing to go to temporary destination group~$b$ that
are delivered during step~$s$.

Consider an arbitrary node~$b\in D^+$.
The expected value of $Y_b^2$ depends on permutation
$\pi'$. Since we are computing a lower bound to $Y_b^2$, the worst case is
when all packets in $D_b^1$ originate at different groups. Indeed, if two packets
in $D_b^1$ belong to the same $S_a^1$, we already know that they have
chosen two different colors during step~$s$, and the expectation of $Y_b^2$ is larger.
A formal proof of this intuitive claim can be
given, though it's omitted for the sake of brevity.
Assuming that random variable $Y_b^1$ is
not far from expectation as in Equation~\ref{eqn:azumaY},
we can bound the expectation of $Y_b^2$,
\begin{align}
\nonumber
\E[Y_b^2] & = |D_b^1|\left(1-\frac{1}{g}\right)^{|D_b^1|-1}\ge\\
\nonumber & \ge (1-\delta)^2|D_b|e^{-|D_b|/g}\left(1-\frac{1}{g}\right)^{|D_b^1|-1}\ge\\
& \ge (1-\delta)^2|D_b| e^{-|D_b|/g}e^{-|D_b^1|/g}\ge (1-\delta)^2|D_b| e^{-2d_s/g}.
\label{eqn:lowerYtilde}
\end{align}
Just as before, also $Y_b^2$ is not far from its expectation.
Martingale theory can be used again to show that
\begin{equation}
\PR\left[\left|Y_b^2-\E[Y_b^2]\right|\ge \delta\E[Y_b^2]\right]
\le 2e^{-\frac{\delta^2 \E[Y_b^2]^2}{2\sum(2)^2}}\le
2e^{-\frac{\delta^2 (1-\delta)^4 g^{5/6}}{8e^2}}.
\label{eqn:azumaYtilde}
\end{equation}
Similarly, by using the same technique that has been used to bound random variable $Y_b^1$,
for every node $a\in S^+$ we can show that
\begin{align}
\nonumber
\E[X_a^2] & \ge (1-\delta)|S_a^1|\left(1-\frac{1}{g}\right)^{|S_a^1|-1}\ge (1-\delta)|S_a^1| e^{-|S_a^1|/g} \ge \\
\nonumber & \ge (1-\delta)^2|S_a|e^{-|S_a|/g} e^{-|S_a^1|/g} \ge\\
& \ge (1-\delta)^2|S_a| e^{-2d_s/g},
\label{eqn:lowerXtilde}
\end{align}
and that $X_a^2$ is not far from its expectation
\begin{equation}
\PR\left[\left|X_a^2-\E[X_a^2]\right|\ge \delta\E[X_a^2]\right]
\le 2e^{-\frac{\delta^2 \E[X_a^2]^2}{2\sum(2)^2}}\le
2e^{-\frac{\delta^2 (1-\delta)^4 g^{5/6}}{8e^2}}.
\label{eqn:azumaXtilde}
\end{equation}

By Equations~\ref{eqn:lowerX}, \ref{eqn:azumaX}, \ref{eqn:lowerP}, \ref{eqn:azumaP}, \ref{eqn:lowerY}, \ref{eqn:azumaY}, \ref{eqn:lowerYtilde}, \ref{eqn:azumaYtilde}, \ref{eqn:lowerXtilde}, \ref{eqn:azumaXtilde}, and
by the union bound,
the number of packets successfully delivered
in step~$s$ can be bounded as follows: For every $\delta>0$,
\begin{align}
\label{eqn:lowerXfinale}
X_a^2 & \ge (1-\delta)^3|S_a| e^{-2d_s/g}\\
\label{eqn:lowerYfinale}
Y_b^2 & \ge (1-\delta)^3|D_b| e^{-2d_s/g}
\end{align}
for every $a\in S^+$ and $b\in D^+$, with probability at least
\begin{equation}
1-9ge^{-\frac{\delta^2 (1-\delta)^4 g^{5/6}}{8e^2}}.
\label{eqn:azumafinale}
\end{equation}

Now, we divide our analysis into two phases. Phase~1 is composed of a constant
number of steps and, with high probability, reduces the maximum degree of the conflict
graph from $d_1$ to $gx$ or less,
where $0\le x<1$ is any fixed constant.
Phase~2 follows and reduces the maximum degree of the conflict graph to $g^{5/6}$ or less
in $O(\log\log n)$ steps with high probability.

Let us start from Phase~1. For every step~$s$ during Phase~1,
$gx\le d_s\le g$. We show that a constant number of
steps is enough to make $d_s$ fall below $gx$ with high probability.
For all $a\in S^+$, let us refer to a step such that
\begin{equation}
X_a^2\ge\frac{|S_a| e^{-2}}{2}
\end{equation}
as a \emph{lucky} step for group~$a$.
By Equation~\ref{eqn:lowerXfinale} and~\ref{eqn:azumafinale}, where we fix
$\delta$ such that $(1-\delta)^3=1/2$, step~$s$ is lucky for every
group~$a\in S^+$ with probability at least
\begin{equation*}
1-9ge^{-\alpha |S_a|}\ge 1-9ge^{-\alpha g^{5/6}},
\label{eqn:azuma}
\end{equation*}
where $\alpha$ is a positive constant.
Therefore, the number of packets that remain after step~$s$ in group~$a\in S^+$ is
\begin{equation}
|S_a|-X_a^2\le |S_a|-\frac{|S_a| e^{-2}}{2}
\le d_s\left(1-\frac{e^{-2}}{2}\right)	
\end{equation}
with high probability.
Note the same bound can be shown for sets~$|D_{b}^1|$, $b\in D^+$, with
exactly the same analysis (where an analogous notion of lucky step refers to a step such that the degree of group~$b\in D$ reduces by $|D_{b}^{s,1}|e^{-2}/2$ at least). Therefore,
after
\begin{equation*}
y:=\left\lceil\frac{\log x}{\log (1-e^{-2}/2)}\right\rceil
\end{equation*}
lucky steps for all the groups the maximum degree of the conflict graph reduces
to $gx$ or less.
By the union bound,
this happens
within the very first $y$ steps
with probability at least
\begin{equation*}
1-9yge^{-\alpha g^{5/6}},
\end{equation*}
That is, Phase~1 completes in a constant number of steps
with high probability.

We are now at a generic step~$s$ in Phase~2.
Our goal is to reduce the degree of the graph of conflicts to $g^{5/6}$.
Let $\lambda_s=d_s/g$. We can assume that $g^{-1/6}\le\lambda_s<x$,
and when $\lambda_s$ falls below $g^{-1/6}$ we are done.
This time, let's refer to a step during which at least
$(1-\lambda_{s})|S_a| e^{-2\lambda_s}$ packets in group~$a\in S^+$ are delivered as a
\emph{lucky} step for group~$a$.
By Equation~\ref{eqn:lowerXfinale} and~\ref{eqn:azumafinale}, where we take
$\delta_s=\lambda_s/3$ (in such a way that $(1-\delta_s)^3\ge (1-\lambda_s)$),
step~$s$ is lucky for every group~$a\in S^+$ with probability at least
\begin{equation*}
1-9yge^{-\beta g^{1/2}},
\end{equation*}
where $\beta$ is a positive constant, 
since $|S_a|\lambda_{s}^2\ge g^{5/6}(g^{-1/6})^2=g^{1/2}$.
So, the number of packets that remain in group~$a\in S^+$ after step~$s$ is
\begin{equation*}
|S_a|-X_a^2\le |S_a|-(1-\lambda_{s})|S_a| e^{-2\lambda_s}\le
d_s\left[1-(1-\lambda_{s}) e^{-2\lambda_s}\right]
\end{equation*}
with high probability. A similar result can be shown
for any group~$b\in D$ such that $|D_b|>g^{5/6}$ with exactly the same analysis.
By the union bound, at the end of step~$s$ the degree of the conflict graph is at most
\begin{equation*}
d_s\left[1-(1-\lambda_{s}) e^{-2\lambda_s}\right]
\end{equation*}
with high probability.
Now, assuming a sequence of lucky steps, we can set up the following recurrence,
\begin{align*}
\lambda_{s+1} & \le \lambda_s \left[1-(1-\lambda_s) e^{-2\lambda_s}\right]\le
\lambda_s\left[1-(1-\lambda_s)(1-2\lambda_s)\right]=\\
& = \lambda_s\left[1-1+3\lambda_s-2\lambda_s^2\right]\le 3\lambda_s^2.
\end{align*}
Therefore,
\begin{equation*}
\lambda_{s}\le 3\lambda_{s-1}^2\le 3\left(3\lambda_{s-2}^2\right)^2\le \dotsb \le
3^{2^{s-y-1}}\lambda_{y+1}^{2^{s-y-1}}.
\end{equation*}
That is,
\begin{equation*}
\log_3\lambda_{s}\le \log_3\left(3^{2^{s-y-1}}\lambda_{y+1}^{2^{s-y-1}}\right)=
2^{s-y-1}\left( 1+\log_3 \lambda_{y+1} \right).
\end{equation*}
Since our first goal is to have $\lambda_s\le g^{-1/6}$, we should find $\bar{s}$ such that
\begin{equation*}
\log_3 \lambda_{\bar{s}}\le -\frac{\log_3 g}{6}.
\end{equation*}
We can get this by taking $\bar{s}$ such that
\begin{equation*}
2^{\bar{s}-y-1}\left( 1+\log_3 \lambda_{y+1} \right)\le -\frac{\log_3 g}{6}.
\end{equation*}
If we choose the arbitrary constant $x$ of Phase~1 to be strictly smaller  than $1/3$, we obtain
that $1+\log_3 \lambda_{y+1}$ is negative, and the above equation comes down to
$\bar{s}=O(\log\log g)$.
Therefore, by the union bound over the $\bar{s}-y-1$ steps of
Phase~2, the whole Phase~2 is made of lucky steps for all the groups in $S^+$ and $D^+$ with
probability at least
\begin{equation*}
1-9(\bar{s}-y-1)ge^{-(\alpha+\beta)g^{\frac{1}{2}}} 
=1-O\left(ge^{-(\alpha+\beta)g^{\frac{1}{2}}}\log\log g
\right).
\end{equation*}

We have shown that, after $\bar{s}=O(\log\log n)$ steps, the maximum degree of
the conflict graph~$G_{\pi}$ is at most $g^{5/6}$ with high probability.
This is enough to get the claim of our theorem
by combining Phase~1 and Phase~2, and then using Lemma~\ref{lem:phase3}.
\end{proof}

We remark that all transmissions occurring during slots~3 and~4
are just acks requiring only ``empty'' messages providing only headers but without payload.
When packets are very long, it may be more efficient to divide the 5 slots into 2 ``short'' slots
and only 3 ``long'' slots, hence profiting from the homogenity of the operations
within a same slot in our routing algorithm.

Note an important property of our algorithm:
processor~$i$ requires enough memory to store at most three packets: one is the original packet~$p_i$,
the second is the
packet whose destination is processor~$i$, and the third
is a copy of another packet as received from group~$\Delta(i)$.
However, if we can assume that packet $p_i$ exits
the network the slot after $p_i$ got to its destination $\pi(i)$,
then the requirement on the internal capacity of processors drops
to only $2$ packets.
Similarly, if we can assume that the input packets are stored on an external feeding line,
then the internal storage requirement drops to $1$.

\subsection{The General Case}

Let start from the case when $d>g$. A natural approach to solve the problem is to perform
two stages: Stage~1 routes the packets until the degree of the conflict graph
is at most $g$; then Stage~2 uses the randomized algorithm described in the previous
section to route the remaining packets in $O(\log\log g)$ slots. Since at most
$g$ packets can be moved without conflicts from each group in each slot, $(d-g)/g$ is a simple
lower bound to the number of slots used in the first of the two above mentioned stages.
In the following, we will show that we are only a constant factor
far from the lower bound, and that we can precisely indicate this factor.

Consider a group~$a\in{\mathbb{N}}_g$.
From this group, there are $d>g$ packets
willing to go to destination. If we let every packet choose
a random destination group and try to reach that group, when $d$ is large (it is
enough that $d=\Omega(g\log g)$) every coupler will have a conflict with high probability
and no packet is delivered. Clearly, this is not what we like to happen. So, the idea for the
first stage of the algorithm is a small modification of the randomized algorithm:
Before participating to the step, every processor with a packet tosses a coin
that says 'yes' with probability~$p$. Only those processors that get a 'yes' are allowed to
participate and send their packet.

In the first step, it is best to choose $p$ equal to $g/d$, in such a way
that $g$ packets are sent on expectation.
This value maximizes the expected number of conflict-less
communications, and thus the number of packets that survive slot~1 and slot~2.
Later on, $p$ has to
be iteratively reduced using a fixed law according to the expected reduction of the number
of packets left in each group.
When at most $g$ packets are left in each group with high
probability, then
we can set $p$ to one, and so proceed with the same algorithm we propose for the case when
$d=g$.

To understand what is the most efficient law, it is important
to understand what is the expected number of packets that are delivered in each step
of the algorithm. Informally speaking, our hope is that exactly $g$ packets from each
group participate to every step of the first phase of the algorithm.
Under this assumption, 
we know that approximately $ge^{-1}$ packets of each group will survive the first slot. At the beginning
of the second slot, these packets are somewhat randomly scattered in the network (not
uniformly at random, unfortunately, as we know from the previous section). If everything
goes just like in the first slot, and this is far from being obvious since the destination is
\emph{not random} now and the packets are \emph{not} distributed
uniformly
at random,
we can hope that $g\exp \{-(1+e^{-1})\}$ packets from each group
survive the second slot
as well, and are thus safely delivered. If this is the case, $\exp \{1+e^{-1}\}((d-g)/g)$ steps
are enough to reduce the number of packets from $d$ to $g$ on expectation.
The following theorem shows that, eventually, what happens is exactly
what we can best hope for. Now, we proceed formally. 
\begin{theorem}
\label{thm:generalcase}
Let $c=\exp (1+e^{-1})\approx 3.927$. A $\POPS$ network can
route any permutation in $5c\lceil d/g\rceil+o(d/g)+O(\log\log g)$ slots
with high probability.
\end{theorem}
\begin{proof}
The idea of the algorithm
is to use $\lceil(c+\epsilon(g))(\frac{d}{g}-1)\rceil$ steps, where $\epsilon(g)=o(1)$,
to reduce the maximum degree of the conflict graph to at most $g$
with high probability. Since every step consists of 5 slots, we then get the claim by
Theorem~\ref{thm:fondamentale}.

Every step~$s$, $s=1,\dotsc, \lceil(c+\epsilon(g))(\frac{d}{g}-1)\rceil$, is similar to the standard step
of the randomized routing
algorithm, with the difference that, before choosing its random color during slot~1, every packet
independently tosses a coin and participates to the step with probability
\begin{equation*}
\frac{g}{d-\frac{g(s-1)}{c+\epsilon(g)}}.
\end{equation*}
Our claim is that, at the beginning of step~$s$, $s=1,\dotsc, \lceil(c+\epsilon(g))(\frac{d}{g}-1)\rceil+1$,
the degree of the conflict graph is at
most $d_s:=d-\frac{g(s-1)}{c+\epsilon(g)}$ with high probability.
As a consequence, when
$s=\lceil(c+\epsilon(g))(\frac{d}{g}-1)\rceil+1$, we get $d_s\le g$ as desired.
The claim is certainly true when
$s=1$. Assume it is true at the beginning of
step~$s\le\lceil(c+\epsilon(g))(\frac{d}{g}-1)\rceil$.
We show that it is true at the beginning of step~$s+1$ as well.

Let $S_a$, $a\in S$, be the set of indices of
the packets in group~$a$ that still have to be delivered at the beginning of
step~$s$. Similarly, let $D_b$, $b\in D$, be the set of indices of the
packets in the whole network that still have to be delivered at the beginning of step~$s$ and
that have group~$b$  as temporary destination group. By hypothesis, $|S_a|\le d_s$ and
$|D_b|\le d_s$ for all $a\in S$ and $b\in D$.
Our first goal is to prove that at the beginning of step $s+1$ the degree of the conflict
graph is at most $d_{s+1}$ with high probability.

For every packet~$p_i$ yet to be routed, let random variable $P_i$ be equal to 1 if packet~$p_i$
participates to step~$s$, and 0 otherwise. Random variable $P_a=\sum_{i\in S_a} P_i$
counts the number of packets in group~$a$ that participate to step~$s$. The expectation of $P_a$
can be computed as follows:
\begin{equation*}
\E[P_a]=\sum_{i\in S_a} \E[P_i]=\frac{|S_a|g}{d_s}.
\end{equation*}
And, clearly, $\E[P_a] \leq g$.
Since random variables $P_i$ are independent, the Chernoff bound~\cite{mr95,as00}
(note that in~\cite{mr95} this bound appears in a different yet stronger form) 
is enough to claim that for every $\delta>0$
\begin{equation*}
\Pr\left[ P_a<(1-\delta)\frac{|S_a|g}{d_s}\right]\le e^{-\frac{\delta^2|S_a|g}{2d_s}}\le
e^{-\frac{\delta^2d_{s+1}g}{2d_s}}\le e^{-\frac{\delta^2g}{4}}
\end{equation*}
and
\begin{equation*}
\Pr\left[ P_a>(1+\delta)g\right]\le e^{-\frac{\delta^2|S_a|g}{2d_s}}\le
e^{-\frac{\delta^2d_{s+1}g}{2d_s}}\le e^{-\frac{\delta^2g}{4}}.
\end{equation*}
Let $S'_a$, $a\in S$, be the set of indices of
the packets in group~$a$ that participate to step~$s$.
Random variable $P_a$ is thus equal to $|S'_a|$.
Therefore, for every $\delta>0$
\begin{equation}
(1-\delta)\frac{|S_a|g}{d_s}\le S'_a \le (1+\delta)g
\end{equation}
with probability at least $1-2e^{-\delta^2g/4}$.
Since a similar result holds for every $a\in S$ and
$b\in D$, we also know that for every $\delta>0$
\begin{align}
\label{eqn:S0lower}
& (1-\delta)\frac{|S_a|g}{d_s}\le S'_a \le (1+\delta)g,\\
\label{eqn:D0lower}
& (1-\delta)\frac{|D_b|g}{d_s}\le D'_b \le (1+\delta)g,
\end{align}
hold for every $a\in S$ and $b\in D$, with probability at least
\begin{equation}
\label{eqn:azumaS0}
1-4ge^{-\delta^2g/4},
\end{equation}
by the union bound over the $2g$ nodes of the conflict graph.

Clearly, we have nothing to show about the nodes in the conflict graph that have degree smaller
than or equal to $d_{s+1}$. So, we define sets $S^+\subseteq S$ and
$D^+\subseteq D$, which collect the nodes with degree
larger that $d_{s+1}$, and focus on the nodes in these sets.
Consider an arbitrary group~$a\in S^+$, and assume that the bound in
Equations~\ref{eqn:S0lower} and~\ref{eqn:D0lower} hold for every $a\in S$ and $b\in D$.
Now, we can perform the same analysis as in the proof of Theorem~\ref{thm:fondamentale}.
Similarly to Equation~\ref{eqn:lowerXtilde}, we know that
\begin{equation*}
\E[X_a^2]  \ge (1-\delta)|S_a^1|\left(1-\frac{1}{g}\right)^{|S_a^1|-1}\ge
(1-\delta)|S_a^1| e^{-|S_a^1|/g},
\end{equation*}
with high probability.
In the next equation, we will use the following two facts: $xe^{x/g}\le ye^{y/g}$
whenever $x\le y\le g$, and $xe^{x/g}$ has maximum when $x=g$.
Clearly, $|S_a^1|\le g$ (there
are only $g$ couplers from group~$a$).
So, we get
\begin{align*}
\E[X_a^2] & \ge (1-\delta)|S_a^1| e^{-|S_a^1|/g}\ge\\
& \ge (1-\delta)^2|S'_a| e^{-|S'_a|/g} e^{-|S'_a| e^{-|S'_a|/g}/g}\ge\\
& \ge (1-\delta)^3  \frac{|S_a|g}{d_s}e^{-1}e^{-e^{-1}}.
\end{align*}
with high probability.
By setting $\delta=g^{-1/3}$ in the above equation, with high probability we get
\begin{equation*}
X_a^2 \ge \frac{|S_a|}{d_s}\frac{g}{c+\epsilon(g)},
\end{equation*}
where $c=e^{1+e^{-1}}$ and $\epsilon(g)=o(1)$.
Since $X_a^2$ is the number of packets in group~$a$ that are delivered
to destination during slot~$s$, the degree of group~$a$ in the conflict graph at the
beginning of step~$s+1$ is
\begin{equation*}
|S_a|- X_a^2 \le |S_a|- \frac{|S_a|}{d_s}\frac{g}{c+\epsilon(g)} \le d_s-\frac{g}{c+\epsilon(g)}
= d_{s+1}.
\end{equation*}
The same result can be shown for every $a\in S^+$ and $b\in D^+$. By the union bound over the
$\lceil(c+\epsilon(g))(\frac{d}{g}-1)\rceil$ steps required, and over the $2g$ nodes in the conflict graph,
and by Equation~\ref{eqn:azumaS0} and a corresponding version of Equation~\ref{eqn:azumafinale},
the degree of the conflict graph is reduced below $g$ with probability at least
\begin{equation*}
1-\left(9ge^{-\delta^2 (1-\delta)^4 g^{5/6}/8e^2}+4ge^{-\delta^2g/4}\right).
\end{equation*}
Note that this is $1-o(1)$ as $g$ grows.
\end{proof}

To get a feeling of the performance of our randomized algorithm, we can set
$\epsilon(g)\approx 0.073$ in the proof of the above theorem, in such a way that
$c+\epsilon(g)=4$. The result is claimed in the following corollary.
\begin{corollary}
\label{cor:general}
A $\POPS$ network can route any permutation in $\frac{20d}{g}+O(\log\log g)$ slots with high probability.
\end{corollary}

\section{Experiments}
\label{sect:exp}

Our results in Theorems~\ref{thm:fondamentale} and~\ref{thm:generalcase} are
asymptotic. In principle, it could thus be possible that the
randomized algorithm does not perform well in practice. This is not the case.
Experiments show that it outperforms the algorithm
in~\cite{ds-IEEETPDS03} even on networks as small as a ${\mathrm{POPS}}(2,2)$,
and proves to be exponentially faster when $d$ and $g$ grow.

The algorithm in~\cite{ds-IEEETPDS03} is claimed to run in $\frac{8d}{g}\log^2 g+
\frac{21d}{g}+3\log g+7$ slots. However, the authors make a small mistake when saying
that Leighton's implementation of the odd-even merge sort algorithm is composed of
$\log^2 n$ steps. The actual complexity is only $\frac{\log n(1+\log n)}{2}\approx 2\log^2 g$ steps.
So, the running time of the routing algorithm in~\cite{ds-IEEETPDS03} is
$\frac{4d}{g}\log^2 g+\frac{2d}{g}\log g+\frac{21d}{g}+3\log g+7$ slots, that is smaller,
and this is what we will use in the following.

To perform the experiments, we built a simulator for the POPS network. It is written in C++
and simulates the network at a message level. That is, for every message in the real network,
there is a message in the simulator.
Processors (implemented as instances of a class \texttt{Processor}) locally take decisions about the next step to perform, and couplers (implemented as instances of a class \texttt{Coupler}) locally
propagate messages or stop them in case of conflicts.

Then, we implemented our randomized algorithm in the
simulator, slot by slot. We have been conservative, no theoretical result is taken for granted and
the randomized algorithm is just simulated message by message.
Not surprisingly, slots~3, 4, and 5 prove to be conflict-less, supporting what is proven
in Proposition~\ref{pro:conflictless}. So, whenever a copy survives slots~1 and~2
it reaches its final destination,
and the associated ack successfully gets to the source processor.
Moreover, three buffers in every processor~$i$ (one for packet~$p_i$, one for packet~$p_{\pi^{-1}(i)}$,
and the third for floating copies of other packets) are enough.

\begin{figure*}
\centering\includegraphics{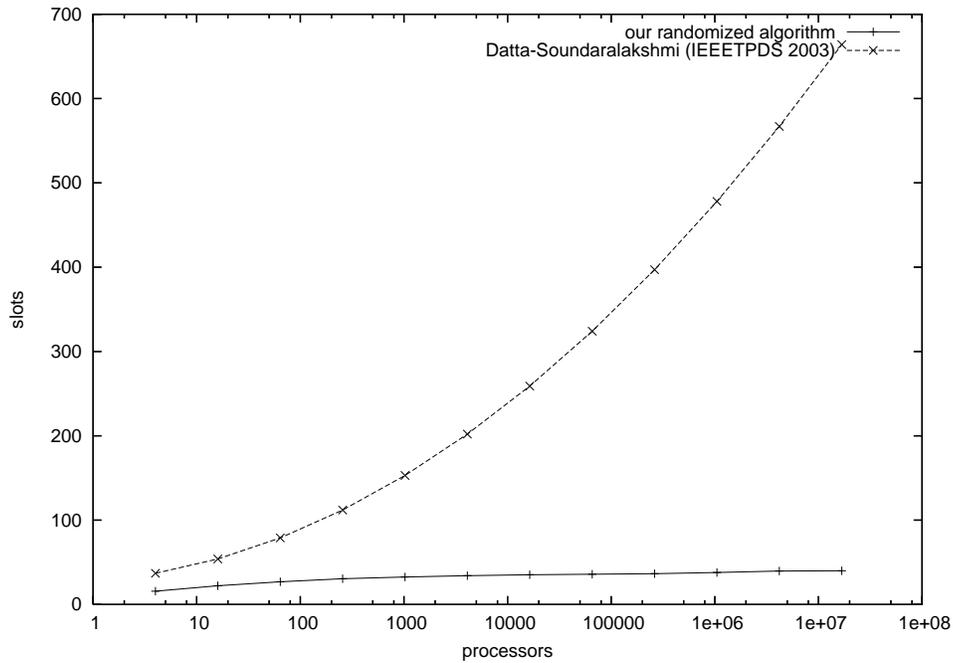}
\caption{Performance of our randomized routing algorithm against the routing
algorithm proposed in~\protect\cite{ds-IEEETPDS03}.
Case when $d=g$. The number of
processors goes from 4 to 16,777,216 (note that axis $x$ is in logscale).}
\label{fig:esperimento-1}
\end{figure*}

In Figure~\ref{fig:esperimento-1},
it is shown the average over a large number of experiments in
the case when $d=g$. The number of processors $n=dg$ goes from 4 to 16,777,216. The
permutation in input is chosen
uniformly
at random from the class of all possible permutations.
It is clear, from the results shown in the figure,
that our algorithm is much faster than the algorithm
in~\cite{ds-IEEETPDS03} even in practice.
Actually, our algorithm outperforms its competitor for all network sizes
hence putting aside any possible concern about the hidden consts.
The performance of our algorithm is so good
that it is actually hard to appreciate it from Figure~\ref{fig:esperimento-1}.
Hence, Table~\ref{tab:esperimenti} shows the exact numerical
results.
\begin{table}
\centering\begin{tabular}{|r||r|r||r|r||r|r|}
\hline
\multicolumn{1}{|c||}{$n$} & \multicolumn{2}{|c||}{$d=g$} &
\multicolumn{2}{|c||}{$d=4g$} & \multicolumn{2}{|c|}{$d=16g$}\\
\hline
& \multicolumn{1}{|c|}{A} & 
\multicolumn{1}{|c||}{B} & \multicolumn{1}{|c|}{A} & 
\multicolumn{1}{|c||}{B} & \multicolumn{1}{|c|}{A} & 
\multicolumn{1}{|c|}{B}\\
\hline
4 & 14.75 & 37 & - & - & - & - \\
\hline
16 & 20.90 & 54 & 71.40 & 118 & - & -\\
\hline
64 & 27.35 & 79 & 82.80 & 177 & 317.90 & 442 \\
\hline
256 & 30.10 & 112 & 87.15 & 268 & 322.45 & 669 \\
\hline
1,024 & 32.50 & 153 & 92.60 & 391 & 343.10 & 1,024 \\
\hline
4,096 & 34.50 & 202 & 94.00 & 546 & 345.60 & 1,507 \\
\hline
16,384 & 35.20 & 259 & 94.95 & 733 & 339.25 & 2,118 \\
\hline
65,536 & 35.55 & 324 & 95.15 & 952 & 336.45 & 2,857 \\
\hline
262,144 & 36.55 & 397 & 95.35 & 1,203 & 334.30 & 3,724 \\
\hline
1,048,576 & 38.25 & 478 & 95.65 & 1,486 & 333.55 & 4,719 \\
\hline
4,194,304 & 39.70 & 567 & 96.25 & 1,801 & 333.05 & 5,842 \\
\hline
16,777,216 & 40.05 & 664 & 97.05 & 2,148 & 333.60 & 7,093 \\
\hline
\end{tabular}
\caption{Number of slots to route a randomly chosen permutation by our randomized algorithm (A) and by the algorithm in
\protect\cite{ds-IEEETPDS03} (B).} 
\label{tab:esperimenti}
\end{table}

\begin{figure*}
\centering\includegraphics{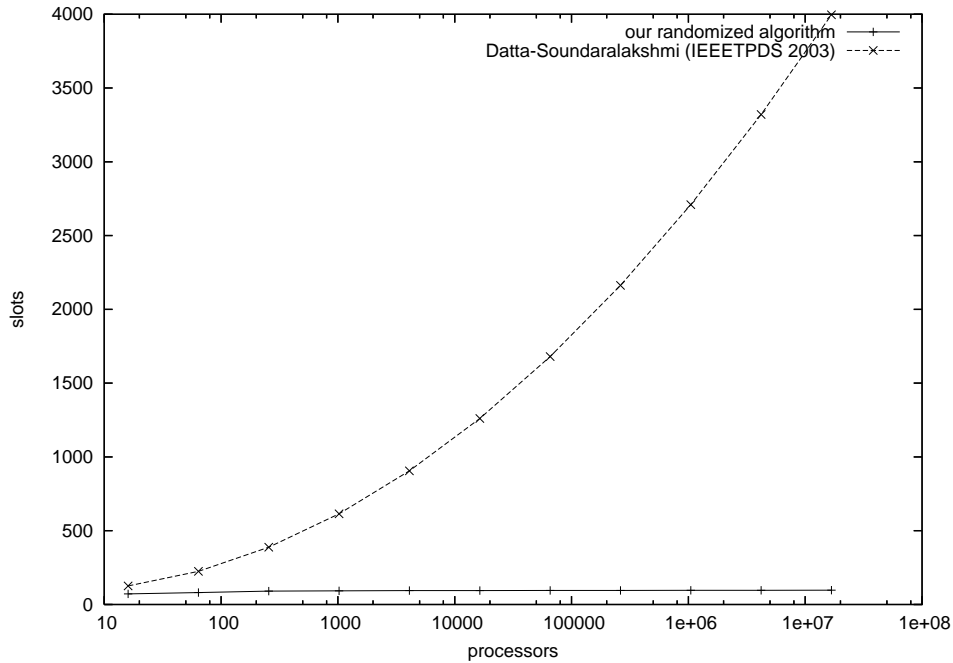}
\caption{Performance of our randomized routing algorithm against the routing
algorithm proposed in~\protect\cite{ds-IEEETPDS03}.
Case when $d=4g$. The number of
processors goes from 16 to 16,777,216 (note that axis $x$ is in logscale).}
\label{fig:esperimento-2}
\end{figure*}

\begin{figure*}
\centering\includegraphics{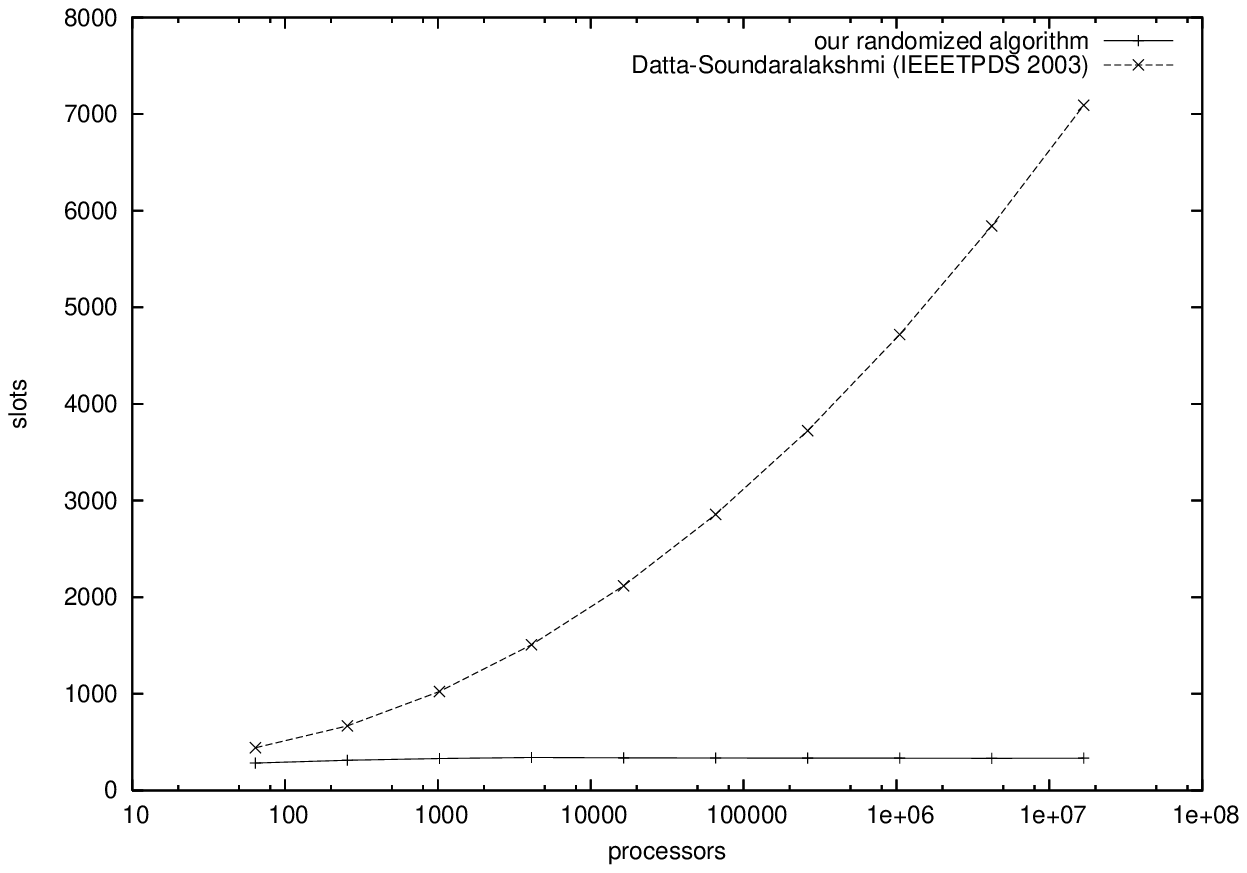}
\caption{Performance of our randomized routing algorithm against the routing
algorithm proposed in~\protect\cite{ds-IEEETPDS03}.
Case when $d=16g$. The number of
processors goes from 64 to 16,777,216 (note that axis $x$ is in logscale).}
\label{fig:esperimento-3}
\end{figure*}

Then, we tested our algorithm on POPS networks with $d$ larger than $g$. We performed
two sets of experiments, one in which $d=4g$ and another in which $d=16g$. In both cases,
the number of processors goes from 4 to 16,777,216. We used
the algorithm as implemented in Corollary~\ref{cor:general}. Therefore, we expect
the routing to take $20\frac{d}{g}+O(\log\log g)$ slot, according to our theoretical results.
In fact, the results that are shown in Table~\ref{tab:esperimenti},
Figure~\ref{fig:esperimento-2}, and Figure~\ref{fig:esperimento-3}
show that the hidden constants are
very small, and that
our algorithm dramatically outperforms the best deterministic algorithm known in the literature for all
network sizes we tested. Finally, Table~\ref{tab:scartp} shows some more details: for each
experiment, we report the average number of steps, the standard deviation, and the worst case
over one hundred runs. Note that the standard deviation is extremely small (smaller than one),
therefore, the performance of our algorithm is almost always very close to expectation.
\begin{table*}
\centering\begin{tabular}{|r||r|r|c||r|r|c||r|r|c|}
\hline
\multicolumn{1}{|c||}{$n$} & \multicolumn{3}{|c||}{$d=g$} &
\multicolumn{3}{|c||}{$d=4g$} & \multicolumn{3}{|c|}{$d=16g$}\\
\hline
& \multicolumn{1}{|c|}{$\mu$} & 
\multicolumn{1}{|c|}{$\sigma$} & \multicolumn{1}{|c||}{max} & \multicolumn{1}{|c|}{$\mu$} & 
\multicolumn{1}{|c|}{$\sigma$} & \multicolumn{1}{|c||}{max} & \multicolumn{1}{|c|}{$\mu$} & 
\multicolumn{1}{|c|}{$\sigma$} & \multicolumn{1}{|c|}{max}\\
\hline
4 & 3.15 & 1.94 & 12 & - & - & - & - & - & -\\
\hline
16 & 4.43 & 1.03 & 8 & 14.33 & 4.22 & 35 & - & - & -\\
\hline
64 & 5.39 & 0.79  & 7 & 16.13 & 2.81 & 27 & 56.88 & 4.52 & 82\\
\hline
256 & 6.10 & 0.57  & 8 & 18.06 & 1.54 & 23 & 62.58 & 3.86 & 81\\
\hline
1,024 & 6.50 & 0.53 & 8 & 18.45 & 0.86 & 20 & 66.26 & 5.16 & 94\\
\hline
4,096 & 6.82 & 0.46 & 8 & 18.81 & 0.64 & 21 & 68.21 & 3.94 & 86\\
\hline
16,384 & 7.04 & 0.20 & 8 & 18.95 & 0.46 & 20 & 67.65 & 1.76 & 73\\
\hline
65,536 & 7.16 & 0.37 & 8 & 19.06 & 0.34 & 20 & 67.12 & 0.89 & 71\\
\hline
262,144 & 7.30 & 0.46 & 8 & 19.09 & 0.29 & 20 & 66.88 & 0.59 & 69\\
\hline
1,048,576 & 7.59 & 0.49 & 8 & 19.15 & 0.36 & 20 & 66.70 & 0.50 & 68\\
\hline
4,194,304 & 7.92 & 0.27 & 8 & 19.21 & 0.41 & 20 & 66.59 & 0.49 & 67\\
\hline
16,777,216 & 8.00 & 0.00 & 8 & 19.41 & 0.49 & 20 & 66.79 & 0.41 & 67\\
\hline
\end{tabular}
\caption{Number of iterations (mean, standard deviation, and worst case over one hundred
runs) to route a randomly chosen permutation by our randomized algorithm.} 
\label{tab:scartp}
\end{table*}

\section{Conclusion}

In this paper, we introduced the fastest algorithms for both deterministic and randomized
on-line permutation routing. Indeed, we have shown that any permutation can be routed on
a $\POPS$ network either with $O(\frac{d}{g}\log g)$ deterministic slots, or, with high probability,
with $5c\lceil d/g\rceil+o(d/g)+O(\log\log g)$ randomized slots, where
$c=\exp (1+e^{-1})\approx 3.927$. The randomized algorithm shows that the POPS network
is one of the fastest permutation networks ever. This can be of practical relevance, since
fast switching is one of the key technologies to deliver the ever-growing amount of bandwidth
needed by modern network applications.

\section*{Acknowledgments}

We are grateful to Alessandro Panconesi for helpful suggestions.

\bibliographystyle{IEEEtran}
\bibliography{r}

\end{document}